 \documentclass{emulateapj}
\usepackage{floatflt,wrapfig}
\usepackage{graphicx}
\usepackage{epsfig}

\usepackage[small]{caption2}

\usepackage{url}
\usepackage{hyperref}

\usepackage{pdfpages}

\usepackage{natbib}
\usepackage{color}
\def\agt{\mathrel{\raise.3ex\hbox{$>$}\mkern-14mu\lower0.6ex\hbox{$\sim$}}}
\def\alt{\mathrel{\raise.3ex\hbox{$<$}\mkern-14mu\lower0.6ex\hbox{$\sim$}}}

\def\be{\begin{equation}}
\def\ee{\end{equation}}
\def\ba{\begin{eqnarray}}
\def\ea{\end{eqnarray}}

\def\mvert{{\vert m\vert}}

\def\msun{M_\odot}
\def\munit{M_{1.4}}
\def\runit{R_{10}}
\def\hz{\,{\rm Hz}\,}
\def\rhob{\rho_b}
\def\rhob14{\rho_{b,14}}
\def\gm{\,{\rm g}}
\def\cm{\,{\rm cm}}
\def\Ts8{T_{S,8}}
\def\sec{{\rm s}}
\def\nuunit{\nu_{500}}

\def\cramp{\vert C_R\vert}
\def\jdot{{\dot J}}
\def\crampjdot{\cramp_\jdot}
\def\ggr{\gamma_{GR}}

\def\yr{\,{\rm y}}
\def\etacc{{\eta_{\rm acc}}}

\def\mdot{\dot M}
\def\crampit{\cramp_{\rm PIT}}
\def\crampitmin{\cramp_{\rm PIT,min}}
\def\domega{\delta\omega}
\def\epsacc{\epsilon_{acc}}
\def\epsacct{\epsilon_{acc,3}}

\def\xivec{{\mbox{\boldmath $\xi$}}}

\def\div{{\mbox{\boldmath $\nabla\cdot$}}}
\def\dotprod{{\mbox{\boldmath $\cdot$}}}

\def\etacore{\eta_{core}}
\def\etacrust{\eta_{crust}}
\def\racc{R_{acc}}

\def\Iunit{{\cal I}_{0.3}}
\def\erg{\,{\rm erg}}
\def\rbunit{r_{b,9}}

\def\gamunit{\gamma_{{\rm acc}, 8}}

\def\ttheta{{\tilde\theta}}

\def\thetap{\theta_+}
\def\thetam{\theta_-}
\def\thetamu{\theta_{\vert\mu\vert}}

\def\cosmu{\vert\mu\vert}

\def\onehalf{{1\over 2}}

\def\kvec{{\mbox{\boldmath $k$}}}
\def\xivec{{\mbox{\boldmath $\xi$}}}

\def\grad{{\mbox{\boldmath $\nabla$}}}
\def\crossprod{{\mbox{\boldmath $\times$}}}
\def\dotprod{{\mbox{\boldmath $\cdot$}}}
\def\zhat{{\mbox{\boldmath $\hat z$}}}
\def\div{\grad\dotprod}

\def\rperp{\varpi}
\def\muvert{\vert\mu\vert}
\def\cone{\cos\theta_1}

\def\ctwo{\cos\theta_2}
\def\sone{\sin\theta_1}
\def\stwo{\sin\theta_2}

\def\nucore{\etacore}
\def\nucrust{\etacrust}
\begin{document}

\title{Nonlinear Development of the R Mode Instability and the Maximum Rotation
Rate of Neutron Stars}
\author{Ruxandra Bondarescu}
\affil{Institute for Theoretical Physics,
University of Zurich, CH-8057, Switzerland}
\email{ruxandra@physik.uzh.ch}
\and
\author{Ira Wasserman\altaffilmark{1}}
\altaffiltext{1}{On leave at KITP UC Santa Barbara}
\affil{Center for Radiophysics and Space Research,
Cornell University, Ithaca, NY 14853}
\email{ira@astro.cornell.edu}

\begin{abstract}
We describe how the nonlinear development of the R mode instability of neutron stars
influences spin up to millisecond periods via accretion. Our arguments are 
based on nearly-resonant interactions of the R mode with pairs of ``daughter modes.''
The amplitude of the R mode saturates at the lowest value
for which parametric instability leads to significant excitation of a
particular pair of daughters. The lower
bound on this limiting amplitude is proportional to the damping rate of the
daughter modes that are excited parametrically. Based on this picture,
we show that if modes damp because of dissipation in a very thin
boundary layer at the crust-core boundary then spin up to frequencies larger
than about 300 Hz does not occur. Within this conventional scenario the
R mode saturates at an amplitude that is too large for angular momentum gain
from accretion to overcome gravitational loss to gravitational radiation. 
We conclude that {\sl lower} dissipation is required for spin up to frequencies much
higher than 300 Hz. We conjecture that 
if the transition from the fluid core to the crystalline crust               
occurs over a distance much longer than $\sim 1$ cm then a sharp viscous boundary
layer fails to form. In this case, damping is due to shear viscosity dissipation
integrated over the entire star; the rate is slower than if a viscous boundary
layer forms. We use statistical arguments and scaling relations to estimate
the lowest parametric instability threshold from first principles.
The resulting saturation amplitudes are low enough to permit
spin up to higher frequencies. Further, we show that 
the requirement that the lowest parametric instability amplitude be small enough to allow continued 
spin up imposes an upper bound to the frequencies that may be attained via accretion that may plausibly
be about 750 Hz.
Within this framework, the R mode is unstable for {\sl all} millisecond
pulsars, 
whether accreting or not.
\end{abstract}

\section{The R Mode Instability Versus the Spin Up Line}
\label{Rmodespin}

The fastest spinning radio pulsar has a rotational frequency $\nu=716$ Hz
\citep{2006Sci...311.1901H}
and 39 have been detected with $\nu>400$ Hz
\citep{2005AJ....129.1993M}. See ATNF Pulsar Catalogue at 
{\url{http://www.atnf.csiro.au/research/pulsar/psrcat/}}.
Moreover, there are 14 pulsars in X ray binaries with inferred 
$\nu>400$ Hz, but none demonstrated convincingly to be faster than 620 Hz
\citep{2012ARA&A..50..609W,2012arXiv1206.2727P}; Chakrabarty has argued
that the population of neutron star spins cuts off sharply at around
730 Hz 
\citep{2005AIPC..797...71C,2008AIPC.1068...67C,2012cosp...39..295C}.
In the standard picture, millisecond pulsars are thought to be spun up
via accretion 
\citep{1982Natur.300..728A}
and the $P-\dot P$ diagram for radiopulsars is consistent with the
idea that accreting neutron stars reach spin equilibrium
\citep[e.g.][]{1997ApJS..113..367B}
in that there appear to be no neutron stars outside the
boundary set by the ``spinup line''
\citep[e.g.][]{1999ApJ...520..696A}
\be
\nu_{eq}=\frac{\omega_s}{2\pi}\sqrt{\frac{GM}{\racc^3}}
\approx\frac{760\hz\omega_s\mdot_9^{3/7}\munit^{5/7}}{\etacc^{3/2}\mu_{26}^{6/7}}
\label{spineq}
\ee
where the accretion radius is $\racc\approx 20\etacc\mu_{26}^{4/7}\mdot_9^{-2/7}\munit^{1/7}$
km, $\mdot=10^{-9}\mdot_9\msun\yr^{-1}$ is the mass accretion rate, $M=1.4\munit\msun$
is the stellar mass,
$\mu=10^{26}\mu_{26}\,{\rm G\,cm^3}$ is the stellar magnetic moment,
and $\omega_s\simeq 1$ and $\etacc\simeq 1$ are parameters that are determined by
the magnetohydrodynamics of disk accretion.
However, there is no particular reason for there to be a spin
frequency cutoff as low as 730 Hz: although there are exceptions,
for most representative equations of state of dense nuclear matter
accretion can spin up a neutron star from $M=1.4\msun$
and $\nu=0$ to $\nu\approx 1000-1500$ Hz before instability ensues
\citep[e.g.][]{1994ApJ...423L.117C}.

The R mode instability can prevent neutron stars from spinning up
to such high frequencies that either dynamical instability or viscosity-driven
secular instability occurs. The R mode instability, 
which is driven slowly by gravitational radiation but stabilized by viscosity,
is reviewed briefly below. However, in the presence of a crust-core boundary layer
the R mode prevents spin-up too efficiently: instability sets 
in at $\nu\approx 300$ Hz (see Eq. (\ref{nucross})), which is too low
to allow for observed frequencies of up to 716 Hz since 
nonlinear effects prevent substantial spin-up
while the star is unstable. We call this {\sl ''The Spin-Up
Problem"}: Phenomenologically, the absence of millisecond pulsars
outside the spin up line up to at least 660 Hz
and the inference that some LMXBs are spinning
faster than 500 Hz suggest that
spin up via slow equilibrium accretion is responsible for the
highest spin frequencies observed, but the R mode 
instability appears to suppress spin up beyond about 300 Hz. 


\section{R Mode Dynamics and the Spin Up Problem}
\label{rmodelow}

The inertial modes of a rotating star may be thought of as
zero frequency ``gauge modes'' of a {\sl non}rotating star
($\delta\rho=-\div(\rho\xivec)=0$)
that acquire frequencies $\vert\omega\vert\leq 2\Omega$ in 
a rotating star to ${\cal O}(\Omega)$
\citep{1978MNRAS.182..423P,1978ApJ...222..281F,1996A&A...311..155L,
2002PhRvD..65b4001S};
the $\ell=\mvert$ R modes are a subset that are axial to ${\cal O}(1)$ for
Newtonian stars, and
have (rotating frame) frequencies $\vert\omega\vert=2\Omega/(\ell+1)$.
(Relativistic modifications have been discussed by 
\cite{PhysRevD.63.024019}.)
For the R modes, ${\cal O}(1)$ displacement fields can be
expressed in terms of a single (magnetic) vector spherical harmonic;
decompositions of the other inertial modes are more complicated
even at ${\cal O}(1)$, generally involving a sum of vector spherical
harmonics up to a maximum $\ell$
\citep[e.g.][]{1999ApJ...521..764L,2000ApJ...529..997Y,
2001ApJ...546.1121Y,PhysRevD.63.024019,PhysRevD.68.124010}.

The R-modes of rotating neutron stars are destabilized by the
emission of gravitational radiation
because their rotating and inertial frame frequencies have
opposite signs, implying that the rotating frame energy
increases as the star radiates energy and angular momentum
in the inertial frame
\citep{1970PhRvL..24..611C,1978ApJ...222..281F,
1998ApJ...502..708A,1998ApJ...502..714F,1998PhRvL..80.4843L,
1998ApJ...501L..89B,1999ApJ...510..846A}.
For the most unstable $\ell_R=m_R=2$ R mode the instability
grows at a rate (numerical coefficients are for the Newtonian
$N=1$ polytrope)
\be
\ggr\approx {\munit\runit^4\nuunit^6\over 2900\sec}
\approx1.6\times 10^{-7}{\munit\runit^4\nuunit^5\,\omega_R}
\label{ggrdef}
\ee
where $M=1.4\munit\msun$ and $R=10\runit$ km are the stellar
mass and radius, and $\nu=500\nuunit\hz$; $\omega_R=2\Omega/3
=4\pi\nu/3$
\citep{1998ApJ...502..708A,1998ApJ...502..714F,1998PhRvL..80.4843L,
1998ApJ...501L..89B,1999ApJ...510..846A}.

Viscous effects (and other forms of dissipation) act against
the instability; the ``CFS stability curve'' in the 
frequency-temperature ($\nu-T$) plane separates stable
and unstable states
\citep[e.g][]{1998PhRvL..80.4843L,1999ApJ...510..846A,2000ApJ...529L..33B,
2002PhRvD..65f3006L,2006PhRvD..73h4001N,2009MNRAS.397.1464H}.
For accreting neutron stars spinning up toward the CFS 
stability curve, balancing accretional heating 
\citep[e.g.][]{2000ApJ...531..988B}
against neutrino cooling implies internal temperature $T\sim
10^8$K
\citep[e.g.][]{2004ARA&A..42..169Y,2008AIPC..983..379Y,2009arXiv0906.1621P}.
At such low temperatures, dissipation in a viscous boundary
layer at the interface between the stellar crust and core
is thought to dominate for the R mode
\citep[e.g.][]{2000ApJ...529L..33B}
implying that the mode first becomes unstable at a spin
frequency
\begin{eqnarray}
\nu_S&\approx& 340\hz(\rhob14/\Ts8)^{2/11} \\ \nonumber
&& \times [10S_R(r_b/0.9R)/\munit\runit]^{4/11}
K_4^{1/11}~,
\label{nucross}
\end{eqnarray}
where $\rho_b=10^{14}\rhob14\,\gm\cm^{-3}$ is the 
density at the crust-core boundary, which is at radius $r_b$, and
$\eta_b=10^4K_4T_8^{-2}\,{\rm cm^2\,s^{-1}}$ 
is the kinematic viscosity at $r_b$,
and $T_S=10^8\Ts8$ K is the temperature.
The quantity $S_R$ measures the imperviousness of the crust to penetration
by the R mode; it depends primarily on the shear modulus of the crust,
but may also be altered by magnetic effects and compressibility
\citep{2001MNRAS.324..917L,2001PhRvD..64d4009M,2003PhRvD..67b4032K,
2006MNRAS.371.1311G}.
\cite{2001MNRAS.324..917L} estimate
that for the (most unstable) R-mode 
$S_R\approx 0.1c_{t,8}/\runit\nuunit$ where 
$c_t=10^8c_{t,8}\cm\sec^{-1}$ is the speed of
crustal shear waves.
The various input parameters are somewhat uncertain. For example
recent 
calculations of the shear viscosity in the core of a neutron
star with improved treatment of dynamical screening
change 
$\eta_b$ by a factor of a few and also alter
its temperature scaling compared to ``traditional'' expressions
\citep{1990ApJ...363..603C,2005NuPhA.763..212A,2008PhRvD..78f3006S}.
While these refinements alter Eq. (\ref{nucross}) slightly,
the 
weak viscosity dependence, $K_4^{1/11}$, still implies 
that $\nu_S$ is well below 716 Hz.

The small value of $\nu_S$ would not be problematic if spinup were to
continue largely unabated within the unstable regime. However, 
detailed nonlinear three mode evolutions using representative
input physics do not support this: the stellar frequency changes
little 
\citep{2007PhRvD..76f4019B}.

%

Two basic principles emerged from our work on multimode
\citep{2002PhRvD..65b4001S,2004PhRvD..70l4017B,2005PhRvD..71f4029B}
and three mode 
\citep{2007PhRvD..76f4019B,2009PhRvD..79j4003B}
nonlinear models for saturation of the R mode
instability:
\begin{enumerate}
\item The R mode amplitude does not grow beyond the first or second
lowest parametric instability threshold amplitude $\crampit$
for interactions with a pair
of daughter modes. $\crampit$
depends on the detuning $\delta\omega=\omega_R-\omega_2-\omega_3$
between the R mode ($\omega_R$) and daughter ($\omega_{2,3}$) 
frequencies, the damping rates of the daughters ($\gamma_{2,3}$)
and the three mode coupling $\kappa$,
\begin{eqnarray}
\crampit^2&=&{\gamma_2\gamma_3\over 4\kappa^2\omega_2\omega_3}
\left[1+\left({\delta\omega\over \gamma_2+\gamma_3}\right)^2\right] \\ \nonumber
&\equiv&\frac{9}{4(\kappa_D\Omega)^2}\left[\gamma_D^2+\frac{(\delta\omega)^2}{4}
\right]~.
\label{crpit}
\end{eqnarray}
Parity and triangle selection rules for the interactions require
that the principal mode numbers of the daughters satisfy the constraint
$n_3=n_2\pm 1$, and for large $n_i$ we expect the viscous damping rates
of the daughter modes to have similar values 
$\gamma_i \approx \gamma_D$ ($i=2,3$); we have also defined $4\omega_2\omega_3\kappa^2=\omega_R^2
\kappa_D^2=4\Omega^2\kappa_D^2/9$.
\item Dissipation of the multitude of daughter modes heats the star
a rate 
\be
H_R=2MR^2\Omega^2\ggr\cramp^2~.
\label{HR}
\ee
Heating proceeds until
balanced by cooling, whereupon evolution tends to settle onto
curves in the $\nu-T$ plane where thermal balance is maintained.
\end{enumerate}
The first basic principle
 is a consequence of the relatively sparse couplings of the
R mode to the sea of daughters
\citep{2002PhRvD..65b4001S,2004PhRvD..70l4017B,2005PhRvD..71f4029B}
and the second merely says that once a steady cascade is set up the
rate at which the R mode sends energy down to the sea equals the
rate of linear growth of its (rotating frame) energy. We stress
that these two principles are based on the {\sl physics of mode coupling}.
Conclusions based on them are more realistic than those based on {\sl ad hoc}
prescriptions for nonlinear truncation of the growth of the R mode
amplitude.

These two principles lead to generic evolution in the $\nu-T$ plane.
The star spins up stably via accretion until it intersects the
stability boundary at $\nu_S$ and $T_S$.The
R mode amplitude then grows rapidly (Eq. [\ref{ggrdef}]) and
reaches $\crampit$ almost immediately. For reasonable parameters,
the R mode heating quickly dominates over accretional heating,
and the star heats up to $T>T_S$. 
Because the cooling, which is
dominated by Cooper pair formation at 
$T_8\simeq 1$
\citep[e.g.][]{1976ApJ...205..541F,1999A&A...343..650Y,
2008PhRvC..77f5808K,2009arXiv0906.1621P}, 
accelerates rapidly,  
fast heating of the star halts eventually. 
Subsequently, the star evolves relatively slowly along 
a track where heating
and cooling are in balance. Whether the star ascends the
curve to higher spin frequency or simply descends to lower
spin frequency depends on whether the spindown due to
gravitational radiation emission is faster or slower than
accretional spinup when heating and cooling first balance.
If the R mode amplitude at this point is large enough, the
star will simply spin down toward the stability curve,
intersecting at $\nu$ slightly below $\nu_S$; otherwise,
the star spins up until gravitational radiation spindown
balances accretional spinup. But even in the latter case,
\cite{2007PhRvD..76f4019B}
found that when damping in a shearing boundary layer dominates
the dissipation
evolutionary
tracks never wander very far from $\nu_S$. Moreover,
once accretion ceases, the star spins down along the curve
where heating and cooling balance, so the end point is virtually
the same as if there were no spin-up in the unstable regime.

Thus, we have two aspects of the {\sl Spin Up Problem}:
\begin{enumerate}
\item The star crosses into the unstable regime at a spin
frequency of about 300 Hz.
\item Saturation of the R mode instability prevents spin up
to higher frequency.
\end{enumerate}

%
%
\begin{figure}
 \epsfxsize = 200pt
 \epsfbox{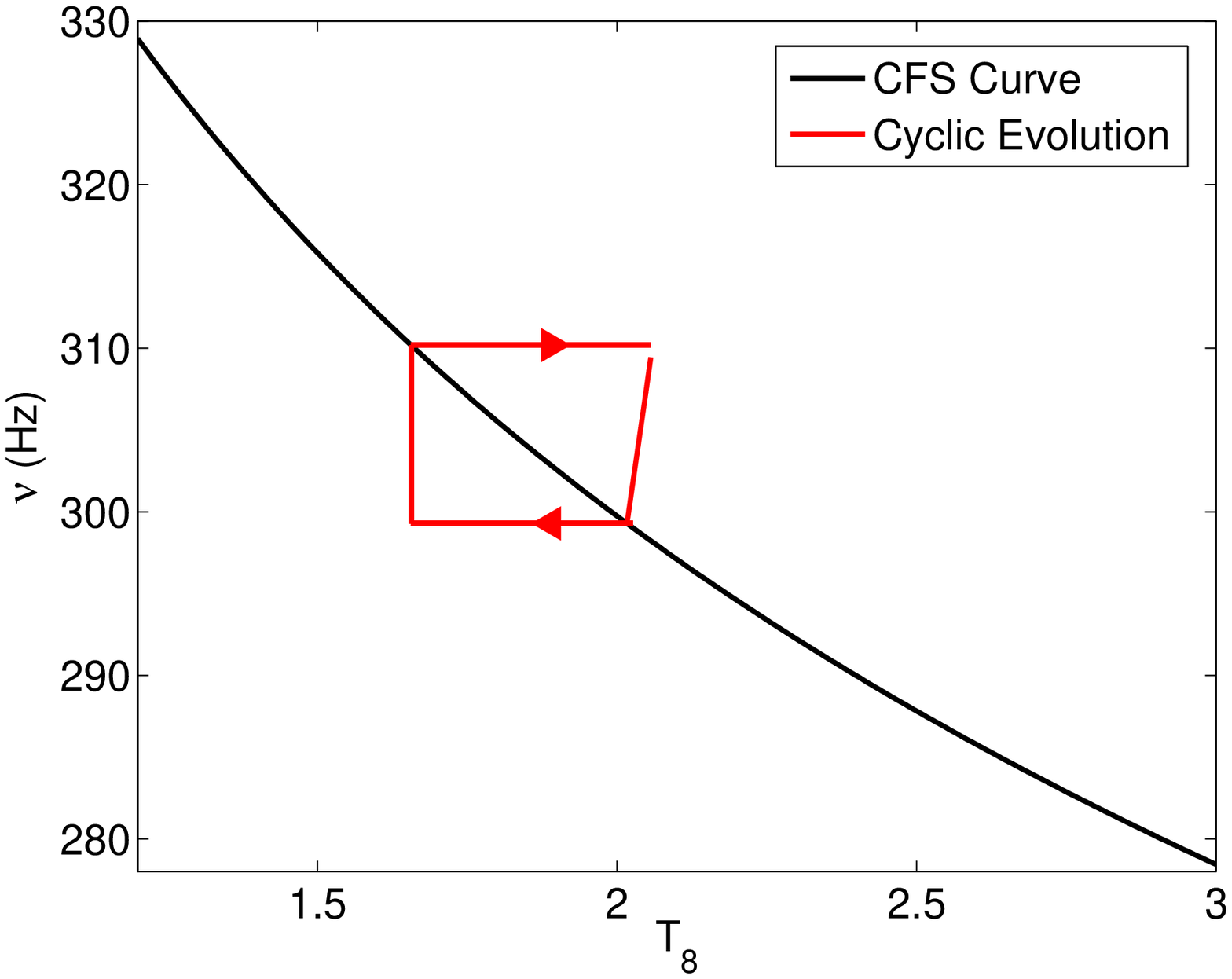}
 \epsfxsize = 200pt
 \epsfbox{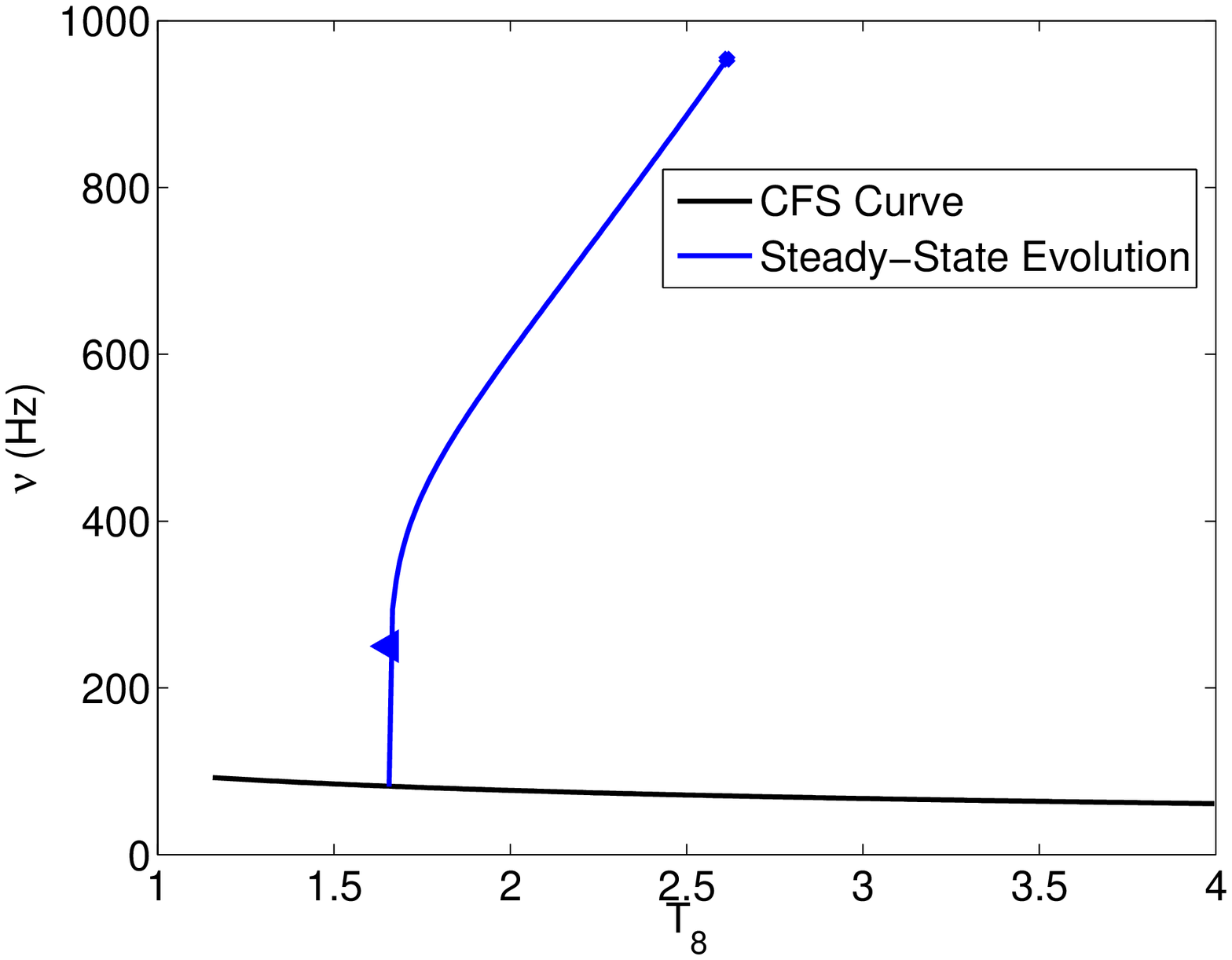}
 \caption{Schematic $\nu-T$ evolutions are shown when the dissipation is dominated by (a) boundary layer viscosity (b) shear viscosity. It can be seen that 
 in the latter case the star spins to much higher frequencies. The final spin frequency where the accretion torque is balanced by gravitational emission is given by Eq.(\ref{nulimitcool}). The CFS instability curve occurs when the gravitational driving equals the viscous damping
  of the R-mode. }
 \label{fig1}
 \end{figure}
The physical reasons that the  evolution is constrained so tightly
can be understood from considering
three different characteristic R mode amplitudes:
\begin{enumerate} 
\item From Eq. (\ref{crpit}), the lowest parametric instability threshold is
$\crampit\gtrsim 3\gamma_D/2\Omega$.  For damping in a shearing boundary layer 
this inequality implies
\begin{eqnarray}
\crampit\gtrsim{3\gamma_D\over 2\kappa_D\Omega}&\approx&\frac{3S_D^2\ell(dE_D/dr)_b}{4\kappa_DE_D} \\ \nonumber
&\approx&\frac{1.3\times 10^{-6}S_D^2K_4^{1/2}(RdE_D/dr)_b}{\kappa_DT_8\nuunit^{1/2}\runit E_D}~.
\label{crampitval}
\end{eqnarray}
Here
$S_D<1$ is the fractional velocity jump across the crust-core boundary for daughter mode $D$, and
$\ell=(\eta_b/\Omega)^{1/2}$ is the boundary layer thickness. 
At principal mode numbers $n_D\gtrsim\omega R/c_t\approx 30\nuunit R_{10}/c_{t,8}$, where
$c_t$ is the transverse shear mode speed in the crust, we expect $S_D\approx 1$; for lower $n_D$,
$S_D<1$. The fractional velocity jump for the $n=3$ R-mode is $S_R\approx 0.1$.
The lowest $\crampit$ arises from modes with $\delta\omega\lesssim\gamma_D$, which are likeliest at
large $n$.
Explicit evaluation for modes of an 
incompressible star as well as WKB calculations for a compressible star
imply that $(R/E_D)(dE_D/dr)_b$ is independent of $n_D$ for $n_D\gg 1$ (see Appendix \ref{blAppendix}).
For incompressible stars, calculations by \citet{2005PhDT........26B} show that $\vert\kappa_D\vert\lesssim 1$
is insensitive to $n_D$, although larger values are likelier at large $n_D$; moreover, $\kappa_D$ is independent 
of $\Omega$ \citep[see][]{2002PhRvD..65b4001S,2003ApJ...591.1129A}.
Thus, the lower bound in Eq. (\ref{crampitval}) is roughly independent of $n_D$. 
In fact, because the lowest expected $\delta\omega$ decreases
with $n$ while $\gamma_D$ is roughly independent of $n_D$,
ultimately $\crampit\simeq 3\gamma_D/2\kappa_D\Omega$ is the lowest parametric instability
threshold for damping in a shearing boundary
layer.

%
\item Gravitational radiation spins the star down at a rate  $-\dot J_{\rm GR}
=6MR^2\Omega\ggr\cramp^2$. If $-\dot J_{\rm GR} < \dot{J}_{acc}$, the star spins up until $\dot J_{\rm GR} = -\dot{J}_{acc}$. Otherwise, if 
$-\dot J_{GR} > \dot{J}_{acc}$, the star will spin down and re-enter the region in which the R-mode is stable.
In spin equilibrium, $J=I\Omega_{eq}\propto
IM^{5/7}\mu^{-6/7}$, where $\Omega_{eq}=2\pi\nu_{eq}$ and $\nu_{eq}$ is given by Eq.~\ref{spineq}.
As accretion proceeds, the magnetic moment decreases and $\nu_{eq}$ increases 
\citep[e.g.][]{1989Natur.342..656S,2006MNRAS.366..137Z}.
A good approximation is 
$\mu\propto(\Delta M)^{-\beta}$, where $\Delta M$ is the total mass
accreted,
and so $J\propto IM^{5/7}(\Delta M)^{6\beta/7}$; \cite{1989Natur.342..656S}
originally suggested $\beta=1$, but \cite{2006MNRAS.366..137Z}
advocate $\beta=7/4$ until $\mu$ ``bottoms out'' at $\mu_{26}\simeq 1$
\citep[see also][]{2011A&A...526A..88W}. In general, the accretion torque is defined to be
$\dot{J}_{acc} = {\dot M}dJ/dM$, which can be written as $\dot J_{acc}/J=\sigma_J\dot M/M$,
where $\sigma_J=6\beta M/7\Delta M+5/7+d\ln I/d\ln M$; spin up is faster
before $\mu$ bottoms out and $\beta\to 0$ and slows as mass accretes and
$\nu$ increases. As a simple model, we adopt $\dot J_{acc}/J=\gamma_{\rm acc}(\nu_0/\nu)^{2s}$;
for numerical estimates, we take $\gamma_{\rm acc} (\nu_0/\nu)^{2s}=10^{-8}\,{\rm y}^{-1}
\,\gamunit\nuunit^{-2s}$, where $s\approx 1/3$ and $s\approx 0.7$ respectively 
before and after $\mu$ bottoms out. The parameter $\gamunit$ is different for each
accreting neutron star.
With this simplified model, $\dot J_{\rm GR}=-\dot J_{acc}$
at an R mode amplitude
\be
\crampjdot\approx\frac{2.1\times 10^{-7}(\Iunit\gamunit)^{1/2}}
{\nuunit^{3+s}\munit^{1/2}\runit^2}~,
\label{spinbalance}
\ee
where the moment of inertia of the star is $I=0.3\Iunit MR^2$.
For numerical estimates, we shall use $s=1/3$, since most of the spin up occurs in
this regime. Comparing Eq. (\ref{spinbalance}) with Eq. (\ref{crampitval}) we see that 
$\crampjdot\lesssim 0.1\crampit$, which means that $\dot J_{GR}>-\dot J_{acc}$,
and spin up is prevented.
\item The amplitude at which heating by the R mode balances heating
via accretion, $H_{acc}=\epsacc\mdot c^2$ with $\epsacc=10^{-3}
\epsacct$ 
\citep{2000ApJ...531..988B}, is
\be
\cramp_H\approx\frac{5.5\times 10^{-8}(\epsacct\mdot_9)^{1/2}}{\nuunit^4\munit\runit^{3}~.
}
\label{heatequal}
\ee
For $\crampit>\cramp_H$, heating by the R mode dominates. Comparing
Eq. (\ref{heatequal}) with Eq. (\ref{crampitval}) implies that heating by the R mode
is more important than accretional heating for damping in a shearing boundary layer.
\item A fourth important amplitude comes from equating gravitational
radiation spindown with $- \dot{J}_B = \eta_{\rm mag}\mu^2\Omega^3/3c^3$, the rate
of pulsar spindown,
\be
\cramp_B\approx\frac{1.5\times 10^{-8}\mu_{26}\eta_{\rm mag}^{1/2}}{\nuunit^{2}\munit\runit^{3}}~.
\label{magspindown}
\ee
This is relevant to the evolution after accretion ceases. If $\crampit>\cramp_B$ then spindown
via gravitational radiation is faster than pulsar spindown.
\end{enumerate}
For damping in a shearing boundary layer, $\crampit > \crampjdot$ 
and so accretion spin-up is limited to about 300 Hz, which is inconsistent with observations 
of pulsars spinning up to $716$ Hz. In Fig. \ref{fig1}, the left panel illustrates a typical evolution sequence
in this case. Because $\crampit>\crampjdot$, the evolutionary track in the $\nu-T$
plane is a rather tight cycle that is confined to a small range of frequencies $\leq\nu_S$, the frequency at
which the mode first becomes unstable, given in Eq. (\ref{nucross}).

For spin up substantially beyond 300 Hz to be possible, $\crampit$ must remain below $\crampjdot$ up to frequencies
well above $\nu_S$. Eq. (\ref{crampitval}) shows that $\crampit\gtrsim 3\gamma_D/2\kappa_D\Omega$, 
but that for damping within
a viscous shearing boundary layer $\gamma_D$ is too large to allow significant spin up. However, Eq. (\ref{crampitval})
also suggests that {\sl lower} $\gamma_D$ would permit prolonged spin up. In \S\ref{success} we examine what happens
if a thin viscous shearing boundary layer does not form near the core-crust boundary, so that $\gamma_D$ is due
to shear viscosity damping distributed over the entire star. In Fig. \ref{fig1}, the right panel illustrates the sort
of evolution that would become possible in this case. As can be seen from the figure, prolonged spin up is possible,
but even in this case there is a maximum attainable spin frequency. We conjecture -- but do not prove ---that if the 
transition from core (super)fluid to crustal solid is gradual enough, a thin viscous shearing boundary layer does
not form. 
%

After accretion ceases, $\crampit$ must 
be small enough that gravitational radiation spin down timescales 
are $\gtrsim 10^9$ years in order for fast spin to be maintained
on spin down timescales characteristic of the fastest millisecond pulsars.
Otherwise, the spun up neutron star would simply spin down too rapidly 
via gravitational radiation, roughly retracing its steps down to the stable
region, leaving a millisecond pulsar with spin frequency $\nu_S\simeq 300\hz$.
Under these circumstances, heating due to the R mode
will be less important than in \cite{2007PhRvD..76f4019B}
during spin up, but may still dominate over accretional heating.
If these conditions can be met,
the star heats modestly after becoming
unstable, but continues to spin up by a significant factor.
After accretion stops, the star cools and spins down within
the unstable regime, but
$\crampit$ is too small to accelerate spin down substantially,
and the spun-up neutron star can become a long-lived millisecond pulsar.
Such a scenario is unfavorable for gravitational radiation detection, but essential
for understanding how pulsars spin up to frequencies $\gtrsim 500$ Hz.

\section{Conditions for a Successful Outcome of Spin-Up}
\label{success}
\def\etad{\eta_{\rm d}}
\def\etamag{\eta_{\rm mag}}

In \S \ref{rmodelow} we demonstrated that the R mode instability frustrates
prolonged spin-up {\sl if} dissipation is due to a viscous shearing boundary
layer at the boundary between the stellar core and crust. The left panel
of Fig. 1 illustrates the problem graphically. If, for some reason, such
a thin viscous boundary layer does not arise, then damping would be 
due to shear viscosity damping distributed over the entire star. This would
result in lower damping rates, and lower $\crampit\gtrsim 3\gamma_D/2\kappa_D\Omega$.

We conjecture that if the transition from fluid core to solid crust occurs
over a radial zone that is considerably thicker than the boundary layer
size, $\ell\simeq(\eta/\Omega)^{1/2}\simeq 1.8K_4^{1/2}\nuunit^{-1/2}T_8^{-1}$
cm,  then a thin viscous boundary layer will not form.
In passing from the fluid core to the solid crust, a mode experiences
a velocity jump $\Delta v$. If the transition from core to crust is abrupt,
then the jump is disontinuous in the inviscid limit. Viscous effects smooth 
the jump so that it occurs continuously;
the smoothing length is $\ell$ which is very small but not zero. Dissipation
within this layer is vigorous, with $\dot E\simeq 4\pi\rho_br_b^2\ell(\Delta v)^2
\times\eta/\ell^2\propto\ell^{-1}$.

Suppose that instead of an abrupt transition, the shear modulus of the star
grows from zero near $r_b$ to its value at the inner edge of the crust
over a radial zone of thickness $\Delta r\gg\ell$. Then we expect the velocity
jump to occur over this relatively extended region. We do not present a 
rigorous calculation of how this happens, since there are many uncertainties,
principally in how the shear modulus grows within the transition region.
The crude toy model developed
in Appendix \ref{toyshear} illustrates the salient features. Fig. \ref{fig:toy} shows
how the displacement field evolves smoothly across the layer in this toy model.
The dissipation associated with this smooth transition is $\dot E\sim 4\pi r_b^2
\Delta r(\Delta v)^2/\times\eta/(\Delta r)^2\propto (\Delta r)^{-1}$. (For
the specific case of the toy model computed in Appendix \ref{toyshear} the constant
of proportionality is about two.) For $\Delta r\gg\ell$, the dissipation rate
associated with a smooth transition layer is much smaller than the dissipation
rate that would arise in a thin boundary layer. For moderate velocity jumps,
such as the relatively small jump associated with the R mode, the extended
transition layer contributes relatively little compared with the energy
dissipation associated with shear viscosity across the entire star.

In the remainder of this section, we assume that dissipation is due to
the distributed effect of shear viscosity. We consider two cases: a ``permeable''
limit where the daughter modes penetrate into the crust, and an ``impermeable''
limit where they do not. We estimate the {\sl lowest value} of $\crampit$ for
each of these cases, taking full account of the two factors in Eq. (\ref{crpit}).
In order to get an estimate, we need scaling relations for $\kappa_D$, $\delta\omega$
and $\gamma_D$ with $n_D$. As was mentioned above, explicit calculations
by \cite{2004PhRvD..70l4017B} \citep[see also][]{2005PhDT........26B} indicate
that $\kappa_D$ does not rise systematically with $n_D$, the principal mode
number of the daughters, although larger values become likelier as $n_D$ 
increases. We use statistical arguments for the expected smallest value
of $\delta\omega$ as a function of $n_D$. We use WKB calculations presented
in \S\ref{ShearViscosityrb}  plus explicit numerical evaluations 
\citep{2004PhRvD..70l4017B,2005PhDT........26B} for $\gamma_D$. The upshot
is that $\delta\omega$ tends to decrease with $n_D$ whereas $\gamma_D$
tends to increase, so there is a minimum value of $\crampit$ at
large values of $n_D$. We shall demonstrate that the minimum occurs at
$n_D\simeq 100$, considerably beyond the ranges computed explicitly even
for the modes of an incompressible star.

\subsection{Permeable Crust}
\label{nonrigid}

Let us consider the non-rigid case first. The WKB calculation detailed in Appendix \ref{ShearViscosityrb} implies
that
\be
\gamma_D\simeq\frac{p_D^2(\etacore-\etacrust)(r_b/R)^2}{R^2\sqrt{1-(r_b/R)^2}}
+\frac{2p_D^3\etacrust}{3R^2}
\label{gammavisc}
\ee
assuming different kinematic viscosities $\etacore$ and $\etacrust$ in the core ($r\leq r_b$) and
crust ($r_b<r\leq 1$), respectively; here $p_D=\sqrt{n_D(n_D+1)-\vert m_D\vert(\vert m_D\vert+1)}
\simeq n_D$. The second term dominates for sufficiently large values of $n_D$, but since 
$\etacrust\ll\etacore$ \citep{2008PhRvD..78f3006S} for practical purposes almost all of the 
dissipation occurs in the core. (For uniform $\etacore=\etacrust$ the second term dominates,
and Eq. (\ref{gammavisc}) agrees with results in \cite{2004PhRvD..70l4017B}.)
Moreover, comparing Eqs. (\ref{transdiss}) and (\ref{gammavisc}), with $\eta_b\sim
\etacore$, we see that dissipation in the bulk of the star dominates over dissipation
in the transition region as long as $p_D\simeq n_D\gtrsim S_D\sqrt{R/\Delta R}=
10S_D\sqrt{R/100\Delta R_t}$. Tentatively, we assume that this inequality holds for
the daughter modes involved in the lowest $\crampit$; we shall see that this is likely to be true.
Thus we adopt 
$\gamma_D=\gamma_0n_D^2$
for estimating the lowest $\crampit$; from the first term in
Eq. (\ref{gammavisc}) with $p_D\simeq n_D$ (as WKB requires)
\be
\frac{\gamma_0}{\Omega}=\frac{\etacore(r_b/R)^2}{\Omega R^2\sqrt{1-(r_b/R)^2}}=
\frac{5.9\times 10^{-12}K_4}{\nuunit T_8^2\runit^2}
\label{gammazero}
\ee
where $\etacore = 10^4K_4T_8^{-2}\,{\rm cm^2\,s^{-1}}=\eta_b$ and we set $r_b=0.9R$.

The other factor in Eq. (\ref{crpit}) is the detuning.
The {\sl minimum} $\delta\omega/\Omega$ up to principal
quantum number $n$ is {\sl expected} to be
approximately $2\sqrt{2}/N(<n)$, where
$N(<n)\approx\frac{1}{6}n^4$ is the number of couplings to the R mode consistent with
selection rules for the transitions
\citep{2005PhDT........26B}. Since we are seeking an estimate of the lowest 
$\crampit$ we substitute this into Eq. (\ref{crpit}) to get
\be
\crampit^2=\frac{9}{4\kappa_D^2}\left(\frac{\gamma_0^2n_D^4}{\Omega^2}+\frac{72}{n_D^8}\right)~;
\label{crampnonrigid}
\ee
recalling that $\kappa_D$ is relatively insensitive to $n_D$ we minimize the quantity
in brackets over $n_D$ and find the lowest value of the threshold at
\be
n_D=1.5\left(\frac{\Omega}{\gamma_0}\right)^{1/6}\approx\frac{110\nuunit^{1/6}T_8^{1/3}\runit^{1/3}}
{K_4^{1/6}}
\label{ndmin}
\ee
where we have used $r_b=0.9R$, and therefore
\be
\crampitmin=\frac{4.2}{\kappa_D}\left(\frac{\gamma_0}{\Omega}\right)^{2/3}
\approx\frac{1.4\times 10^{-7}K_4^{2/3}}{\kappa_D\nuunit^{2/3}T_8^{4/3}\runit^{4/3}}~.
\label{crampitmin}
\ee
Requiring that $\crampit < \crampjdot$ implies that
nonlinear dynamics limits spin up to frequencies
\be
\nu\lesssim\frac{590\hz\kappa_D^{3/8}T_8^{1/2}(\Iunit\gamunit)^{3/16}}
{K_4^{1/4}\munit^{3/16}\runit^{1/4}}~,
\label{numax}
\ee
where we have used $s=1/3$ to obtain the numerical value.
The existence of a maximum spin frequency limit for spin up via accretion is a
generic feature of the dynamics: $\crampitmin$ is determined by a competition
between the decrease of the smallest expected detuning $\domega$ and the 
increase of the dissipation $\gamma_D$ with increasing $n_D$. 

Including
other physical features that we have neglected here will not do away with this
key feature of the dynamics. Two physical features we shall study subsequently
are buoyancy and relativistic corrections. Buoyancy shifts mode frequencies,
but {\sl not} that of the R mode
\citep{1982ApJ...256..717S,2000ApJ...529..997Y}, and also activates the $n\neq\vert m\vert+1$
r modes in the star 
\citep{1982ApJ...256..717S,2000ApJS..129..353Y}.
Relativistic corrections also shift mode frequencies
\citep[e.g.][]{PhysRevD.63.024019,PhysRevD.68.124010},
and may also generate non-axial contributions to the R mode eigenfunction that
permit additional couplings that would be forbidden non-relativistically
\citep[see e.g.][]{PhysRevD.63.024019,PhysRevD.68.124010}. 
Studies that combine buoyancy and relativity are tricky 
\citep[e.g.]{1998MNRAS.293...49K,1999ApJ...520..788K,2007PhRvD..75d3007B,PhysRevD.77.024029}
but detailed calculations seem to support the existence of a mode structure very similar to the
Newtonian case 
\citep{2004CQGra..21.4661L,2001PhRvD..63b4019L,PhysRevD.68.124010,2005MNRAS.363..121P,PhysRevD.71.083001}.
In any event, including both buoyancy and relativistic corrections will not alter the key feature of
the network of interacting modes, namely that there exists a dense set of frequencies bounded
above and below, which permits an increasing number of near resonances as $n_D$ increases.
Moreover, additional couplings may become possible that would be forbidden otherwise,
which could lower the value of $\crampitmin$, thus permitting spin up to larger $\nu$.  
Differential rotation and magnetic fields \citep{2002ApJ...574..908M,Luciano1,Luciano2,Luciano3} and mutual friction 
\citep{Haskell13} may play an important role in limiting the R-mode amplitude.
More work is needed to investigate such effects in detail.

The existence of a maximum frequency dictated by the nonlinear dynamics is a basic 
conclusion of this paper. The actual value of the maximum frequency depends on the 
external variable, $\gamunit$, even though the dependence is weak. Each individual neutron star
has its own value of $\gamunit$, so the maximum spin rate that is attainable is {\sl not} the
same for all neutron stars. We emphasize that our estimate of $\crampitmin$ is the key
to determining the value of the maximum spin reate. The saturation amplitude of the R mode
is not an adjustable parameter, but is determined by the nonlinear hydrodynamics of the
network of interacting modes.

It is reassuring that the value of the maximum frequency in Eq. (\ref{numax}) is
close to 700 Hz, but to go further we need the value of $T_8$ in particular; 
this is determined from balancing heating and neutrino
cooling.
We determine $T$ from the
relationship 
\be 
L_\nu=H_{acc}+H_R,
\ee 
where $L_\nu$ is the neutrino cooling rate.
We assume that cooling is primarily via the Cooper pair process, with
$L_\nu\simeq 10^{33}f_\nu
T_8^8\erg\sec^{-1}$ where $f_\nu\sim 1$ may depend on $M$ and $R$
\citep[e.g.][]{2004A&A...423.1063G,2004ApJS..155..623P,2011PhRvL.106h1101P,2011MNRAS.412L.108S}.  
We evaluate $H_R$ using $\crampitmin$ from Eq. (\ref{crampitmin}); with $H_R=\epsacc{\dot Mc^2}$ we find that
thermal balance implies
\be
f_\nu T_8^8=57\mdot_9\epsacct
+\frac{360\nuunit^{20/3}K_4^{4/3}\munit^2\runit^{10/3}}
{\kappa_D^2T_8^{8/3}}~,
\label{heatingbalance}
\ee
which defines a curve in the $\nu-T$ plane along which the star evolves during
acrretion.
Eq. (\ref{heatingbalance}) shows that $H_{acc}$ dominates at low $\nuunit$ 
(e.g. where the instability
ensues), and $T_8\approx 1.7(\mdot_9\epsacct/f_\nu)^{1/8}$ in this regime;
$H_R$ dominates at large $\nuunit$ (i.e. where the upper spin limit is
fixed) and
\be
T_8\approx\frac{1.7\nuunit^{5/8}K_4^{1/8}\munit^{3/16}\runit^{5/16}}{\kappa_D^{3/16}f_\nu^{3/32}}
~.
\label{TRheat}
\ee
Using Eq. (\ref{TRheat}) in Eq. (\ref{numax}) implies a more precise upper bound 
\be
\nu\lesssim
\frac{950\hz\kappa_D^{9/22}(\Iunit\gamunit)^{3/11}}{f_\nu^{3/44}K_4^{3/11}(\munit\runit)^{3/22}}
\equiv\nu_{max}~.
\label{nulimitcool}
\ee
The full solution of Eq. (\ref{heatingbalance}) would give a slightly lower value.
%

Because $\gamma_D$ is low, spin up begins at a significantly lower frequency than
Eq. (\ref{nucross}), typically $\nu_S\simeq 100-150\hz$, so prolonged spin up via
accretion is required. Throughout much of this evolution, the R mode plays almost
no role 
because of the strong frequency dependences of $H_R/H_{acc}$ and $\crampitmin/\crampjdot$.
Spin up ends either because accretion ceases or
because spin equilibrium $\crampit=\crampjdot$ is achieved.
In the former case, spin up proceeds almost as it would if there were no
R mode instability. 
At its maximum spin frequency, a neutron star is in spin balance, with equal and opposite gravitational
radiation and accretion torques, and remains in that state until accretion ends. Depending on the detailed
evolution, spin equilibrium can occupy a substantial fraction of the time during which a neutron star
accretes. In thermal balance, Eq. (\ref{heatingbalance}) shows that the neutron star's internal 
temperature is an increasing function of frequency, but also depends on $\mdot_9$, which is different for
each accreting neutron star, and $\kappa_D$. Although
we expect {\sl similar} values of $\kappa_D$ for different neutron stars, they need not be
{\sl identical}, because $n_D$ is not the same for all neutron stars affected by the R mode instability. 
Thus, there may be
some variability in internal and effective temperatures for neutron stars in the unstable domain.
Intermittent accretion is unlikely
to affect these conclusions: cooling timescales are $\sim 100-1000$ years
so if the heating rate fluctuates at much shorter timescales the time averaged
heating rate is all that matters.
Similarly, the detailed time dependent dynamical evolution of the R mode proceeds on timescales
that are too short, $\sim 1/\delta\omega\sim 1/\crampit\Omega$, to be important for the secular
evolution of spin and internal temperature; whether there are any observable effects of the dynamics
is beyond the scope of this paper.

Once accretion ends, the fast rotating neutron star cools and spins down. Because
the cooling timescale is short 
compared with the spin
down timescale ($\gtrsim 10^8\gamunit^{-1}$ years) at the end of spin up, the
neutron star first cools at fixed spin frequency. Cooling ends when $H_R$
is balanced by cooling.
Once this point is reached, the neutron star spins down along the curve
given by Eq. (\ref{TRheat}), and 
\be
\crampitmin\approx\frac{6.6\times 10^{-8}K_4^{1/2}f_\nu^{1/8}}{\nuunit^{3/2}\kappa_D^{3/4}
\munit^{1/4}\runit^{7/4}}~.
\label{crampitendgame}
\ee
Slow evolution along this curve is driven by 
spin down: if there were no change in $\nu$ the star would remain at a single point
in the $\nu-T$ plane. The total spin down rate is the sum of contributions from
gravitational radiation and electromagnetic radiation $- \dot{J}=\dot{J}_{GR} + \dot{J}_B$. The spin-down rate at the lowest PIT given by
Eq. (\ref{crampitendgame}) is
\be
-\frac{\dot J}{I\Omega}
\approx\frac{(\nu/\nu_{max})^{11/3}\gamunit}{10^8\,{\rm y}\,\nu_{max,500}^{2/3}}
+\frac{\etamag\mu_{26}^2\nuunit^2}{14.5\times 10^9\,{\rm y}\,\Iunit\munit\runit^2}~,
\label{spindownrate}
\ee
where we have used $\nu_{max,500}=\nu_{max}/500\hz$. 
Spin down ages $t_{sd}=-I\Omega/2\dot J$ for radiopulsars with $\nu\gtrsim 400\hz$ range between
$1.64\times 10^8$ y and $14.3\times 10^{10}$ y; 
see 
\cite{2005AJ....129.1993M}, {\url{http://www.atnf.csiro.au/research/pulsar/psrcat/}}.
If we require a spindown age $\gtrsim 10^9$ years at $\nu_{max}$, Eq. (\ref{spindownrate}) implies
that $\gamunit\lesssim 0.05$. Inserting this into Eq. (\ref{nulimitcool}) lowers $\nu_{max}$,
keeping all other parameters fixed. Since $\nu_{max}\propto(\kappa_D\gamunit^{2/3}/K_4^{2/3})^{9/22}$,
the bound can still be around 750 Hz if 
$\kappa_D/K_4^{2/3}\gtrsim 4.1$; values of $\kappa_D$ this large are unusual but not unheard
of for incompressible stars \citep{2004PhRvD..70l4017B,2005PhDT........26B}, and it is conceivable
that $K_4\lesssim 1$.
This scenario for millisecond pulsar formation
requires that {\sl all} of the fastest spinning pulsars are in the {\sl unstable} domain.
Eq. (\ref{spindownrate}) predicts spin down indices $n=\nu\ddot\nu/\dot\nu^2>3$.
Determinations of $\ddot\nu$ for millisecond pulsars are contaminated by timing noise so there is no 
conclusive evidence against this picture.

\subsection{Impermeable Crust}
\label{rigid}

Calculations for the rigid case follow closely the methodology of \S\ref{nonrigid} but there
is an important difference: because the daughter modes are confined to the core, for practical purposes
$R$ is replaced by $r_b$ in the WKB solutions. We regard this as an extreme limit, and that
more realistically, for the values of $n_D$ we estimate below, the daughter modes penetrate the crust
incompletely with a fractional velocity jump $S_D\lesssim 1$.
In this case, we get a damping rate
\be
\gamma_D=\frac{2p_D^3\etacore}{3r_b^2}
\ee
i.e. we get the second term in Eq. (\ref{gammavisc}) with $R\to r_b$ and $\etacrust\to\etacore$.
There is no need to include the effect of the transition region, since it is already included
(and partly responsible for the stronger scaling with $p_D$).
Instead of Eq. (\ref{crampnonrigid}) we get
\be
\crampit^2=\frac{9}{4\kappa_D^2}\left(\frac{\gamma_0^2n_D^6}{\Omega^2}+\frac{72}{n_D^8}\right)
\label{cramprigid}
\ee
but with (letting $r_b=9\rbunit$ km)
\be
\frac{\gamma_0}{\Omega}=\frac{\etacore}{3\Omega r_b^2}
=\frac{
2.6\times 10^{-12}K_4}{\nuunit T_8^2\rbunit^2}
~.
\ee
Neglecting variations in $\kappa_D$ as before, Eq. (\ref{cramprigid}) is minimize at
\be
n_D=1.4\left(\frac{\Omega}{\gamma_0}\right)^{1/7}\approx\frac{
63\nuunit^{1/7}T_8^{2/7}\rbunit^{2/7}}
{K_4^{1/7}}
\ee
and
\be
\label{crampminrigid}
\crampitmin=\frac{5.3}{\kappa_D}\left(\frac{\gamma_0}{\Omega}\right)^{4/7}
\ee
Following the same procedure as led to Eqs. (\ref{nulimitcool}) and (\ref{crampitendgame})
leads to the final results 
\be
\nu\lesssim\frac{
360\hz\kappa_D^{7/18}(\Iunit\gamunit)^{1/4}\rbunit^{4/9}}{K_4^{2/9}f_\nu^{1/18}\munit^{5/36}\runit^{2/3}}
\equiv\nu_{max}
\label{nulimfinalrigid}
\ee
and
\be
\crampitmin\approx\frac{
4.1\times 10^{-7}K_4^{4/9}f_\nu^{1/9}}{\nuunit^{4/3}\kappa_D^{7/9}\munit^{2/9}
\runit^{2/3}\rbunit^{8/9}}~.
\label{crampminrigidendgame}
\ee
Eq. (\ref{nulimfinalrigid}) requires $\kappa_D\gamunit^{9/14}/K_4^{4/7}\gtrsim 6.6$ for $\nu_{max}\approx 750\hz$,
holding all other paramters fixed. 
Eq. (\ref{crampminrigidendgame}) implies a spin down rate
\be
-\frac{\dot J}{I\Omega}
\approx\frac{(\nu/\nu_{max})^4\gamunit}{10^8\,{\rm y}\,\nu_{max,500}^{2/3}}
+\frac{\etamag\mu_{26}^2\nuunit^2}{14.5\times 10^9\,{\rm y}\,\Iunit\munit\runit^2}~.
\label{spindownraterigid}
\ee
Just as we found for the nonrigid case, we need to cut down the gravitational radiation
contribution in order to be consistent with pulsar data: requiring a spin down timescale
due to gravitational radiation $\gtrsim 10^9$ years near $\nu_{max}$ implies
$\gamunit\lesssim 0.05$, and for $\nu_{max}\simeq 750\hz$ we would then require
$\kappa_D/K_4^{4/7}\gtrsim 45$, which is a more stringent constraint than we found in
\S\ref{nonrigid}.

\section{Conclusions}

Our examination of the nonlinear dynamics of rotational modes of a neutron star suggests that 
in the conventional picture, where modes damp in a thin viscous boundary layer, spin up beyond 
about 300 Hz is not possible: not only is the frequency at which the R mode first destabilizes 
about 300 Hz (see Eq. (\ref{nucross})) but the R mode amplitude saturates at a level large
enough that gravitational radiation spindown prevents significant spin up subsequently
(see Eq. (\ref{crampitval})).
Thus, we consider what happens if a thin shearing boundary layer cannot form. We conjecture
that this may happen if the
transition between core and crust occurs in a region thicker than $\sim 1-2$ cm, and 
justify that assumption partially with the toy model in Appendix \ref{toyshear}.  If a thin
boundary layer does not form,
damping of all modes is dominated by the distributed effects of shear viscosity throughout the star,
which leads naturally to a lower R mode saturation amplitude. 

Using scaling relations found by a combination of exact calculations for 
incompressible stars \citep{2004PhRvD..70l4017B,2005PhDT........26B},
statistical arguments \citep{2005PhDT........26B}
and approximate WKB calculations (Appendix \ref{wkb}) we estimate
the lowest parametric instability threshold $\crampitmin$ analytically for
coupling of the R mode to pairs of daughters.
We find that the daughter modes for which this occurs are at principal mode
quantum $n_D\simeq 100$, typically; this is beyond the range for which
explicit calculations exist, even for incompressible stars. We stress that
the lowest parametric instability threshold sets the amplitude at which
the R mode amplitude saturates during evolution of a network of rotational
modes of a neutron star \citep{2005PhRvD..71f4029B}. Thus, our estimate
of $\crampit$ represents a first principles calculation of the saturation
amplitude. We stress that this is {\sl not} an adjustable parameter, but
rather arises from the nonlinear hydrodynamics. Although it may seem
counterintuitive, when there are many nearly resonant modes, as is the case
for a rotating neutron star, nonlinear effects become important at low
amplitude, and lead to saturation.

With the lower $\crampitmin$ that arises when shear viscosity dominates the damping,
prolonged spin up to frequencies above 500 Hz is possible.
A basic conclusion is that the nonlinear development of the R mode instability naturally gives rise 
to an upper spin frequency limit.
This bound arises from the requirement that $\crampitmin$ be smaller than $\crampjdot$, the amplitude where gravitational radiation
spin down balances accretion spin up. Eq. (\ref{nulimitcool}) and Eq. (\ref{nulimfinalrigid})
provide rough estimates for the maximum spin frequency $\nu_{max}$ that can be attained under the
assumption that the crust is permeable and impermeable to small scale modes, respectively.
It is plausible that $\nu_{max}\simeq 750\hz$, but consistency with observations of millisecond
pulsars requires relatively strong (but not outrageously strong) coupling $\kappa_D$; smaller
values are allowed for the permeable case, which may argue in its favor.
This suggests that nonlinear
interactions among the rotational modes of a neutron star may naturally imply a maximum spin
frequency below what one might expect from dynamical instabilities of the star.
This conclusion is compatible with studies that suggest that LMXBs are not spun up
beyond about 730 Hz \citep{2005AIPC..797...71C,2008AIPC.1068...67C,2012cosp...39..295C}
as well as the fact that the fastest spinning neutron star yet discovered spins at 716 Hz.

A second conclusion of our study is that after accretion ceases, fast spinning millisecond
pulsars cool until they reach a balance between neutrino cooling and heating that results 
from the energy sent to smaller scale modes from the unstable R mode.
The result is slow evolution along a curve in the $\nu-T$ plane, Eq. (\ref{TRheat}).
Spindown timescales are sufficiently long that once spun up a millisecond pulsar ought to
remain close to the upper part of this curve. This means that millisecond pulsars 
remain stuck in
the domain where the R mode is unstable, and are therefore radiating gravitational radiation.
However, the emission rate is very low, and strain amplitudes at Earth are correspondingly low,
$\sim 10^{-26}/D_{kpc}t_{sd,9}$ for a source at $D=D_{kpc}$ kpc with a spin down
time $10^9t_{sd,9}$ years. Although gravitational radiation may dominate the spin down,
because the accretion spin up rate  generally sets torque amplitudes we expect millisecond
pulsars to be near the conventional spin up line but possibly slightly above it.

\appendix
\section{Assorted WKB Results}
\label{wkb}
\subsection{Preliminaries: Coordinates}
\label{prelim}

The Bryan coordinates $x_{1,2}$: for a mode with $\omega=2\Omega
\muvert\equiv 2\Omega\cos\thetamu\leq 2\Omega$ are
\ba
\rperp&=&\sqrt{x^2+y^2}=\sqrt{(1-x_1^2)(1-x_2^2)\over 1-\mu^2}
=\frac{\sin\theta_1\sin\theta_2}{\sqrt{1-\mu^2}}
\nonumber\\
z&=&{x_1x_2\over\muvert}=\frac{\cos\theta_1\cos\theta_2}{\muvert}
\nonumber\\
x_1&\in&[\muvert, 1]~~~~x_2\in[-\muvert,\muvert]\nonumber\\
\theta_1&\equiv&\cos^{-1}(x_1)\in[0,\thetamu]~~~
\theta_2\equiv\cos^{-1}(x_2)\in[\thetamu,\pi-\thetamu]~.
\label{setup}
\ea
We use units in which the radius of the star is $R=1$.
The following are useful definitions: with $\theta_\pm=\theta\pm\thetamu$
\ba
\cos(\theta_2-\theta_1)&=&z\muvert+\rperp\sqrt{1-\mu^2}=r\cos\theta_-
\nonumber\\
\cos(\theta_2+\theta_1)&=&z\muvert-\rperp\sqrt{1-\mu^2}=r\cos\theta_+~.
\label{thetapm}
\ea
For finding mode displacements, we will want derivatives of 
$\theta_{1\,{\rm or}\,2}$ with respect to coordinates. In
cylindrical coordinates Eq. (\ref{setup}) implies
\ba
\sqrt{1-\mu^2}\,d\rperp&=&\cos\theta_1\sin\theta_2d\theta_1
+\sin\theta_1\cos\theta_2d\theta_2
\nonumber\\
\muvert\,dz&=&-\sin\theta_1\cos\theta_2d\theta_1
-\cos\theta_1\sin\theta_2d\theta_2
\nonumber\\
d\theta_1&=&{\cos\theta_1\sin\theta_2d\rperp\sqrt{1-\mu^2}
+\sin\theta_1\cos\theta_2 dz\muvert\over
\cos^2\theta_1\sin^2\theta_2-\sin^2\theta_1\cos^2\theta_2}
\nonumber\\
d\theta_2&=&{\cos\theta_2\sin\theta_1d\rperp\sqrt{1-\mu^2}
+\sin\theta_2\cos\theta_1dz\muvert\over
\cos^2\theta_2\sin^2\theta_1-\sin^2\theta_2\cos^2\theta_1}
\label{thetaderivscyl}
\ea
Using Eqs. (\ref{thetaderivscyl}) 
we find the area element 
\be
dA=\rperp d\rperp dz=
{\sone\stwo(\cone^2-\ctwo^2)d\theta_1d\theta_2\over\cosmu(1-\mu^2)}
={(x_1^2-x_2^2)dx_1dx_2\over\muvert(1-\mu^2)}~.
\label{useful}
\ee
(The integral $\int dA$ over the ranges of $\theta_{1,2}$ 
or $x_{1,2}$ is 2/3.)
The stellar surface $r=1$
is patched together in the following way:
\ba
\theta_1&=&\thetamu~~~{\rm and}~~~\theta_2=\theta\in[\thetamu,\pi-\thetamu]
\nonumber\\
\theta_2&=&\thetamu~~~{\rm and}~~~\theta_1=\theta\in[0,\thetamu]
\nonumber\\
\theta_2&=&\pi-\thetamu~~~{\rm and}~~~\theta_1=\pi-\theta\in[0,\thetamu]
~.
\label{onsurface}
\ea
There are special points
where $x_1^2-x_2^2=0=dA$; at these points, 
$\cone=\pm\ctwo=\cosmu$. 

\subsection{WKB Approximation to Displacements} 

From 
\cite{2003ApJ...591.1129A} \S 3.2 we take the WKB Eulerian enthalpy
perturbation to be
\footnote{However, we use the convention that the mode is
proportional to $\exp(+i\omega t)$; 
\cite{2003ApJ...591.1129A}
employed modes $\propto\exp(-i\omega t)$.}
\be
\Psi\approx{P_{nm}(x_1)P_{nm}(x_2)\exp[i(m\phi+\omega t)]
\over\sqrt{\rho}}\approx
{\cos(p\theta_1+\alpha_1)\cos(p\theta_2+\alpha_1)\exp[i(m\phi+\omega t)]
\over\sqrt{\rho\sin\theta_1\sin\theta_2}}
\label{wkbenth}
\ee
where $\rho=\rho(r)$ is the density profile 
and $p=\sqrt{n(n+1)-m(m+1)}\simeq n$. The first approximation 
assumes that the density scale height is large compared with 
characteristic scales on which $\Psi$ varies. The second approximation
is for the associated Legendre functions, and holds at sufficiently
large values of $p$. The values of the phases $\alpha_i$ depend on the parity
of the mode: based on asymptotic properties of the $P_{nm}(z)$,
\cite{2003ApJ...591.1129A} adopted
$\alpha_1=\alpha_2=-p\pi/2$ or
$=-(p+1)\pi/2$ even or odd parity, respectively, 
but \cite{2010MNRAS.407.1609I}, using a more delicate treatment of 
boundary conditions, argued for $\alpha_1\neq \alpha_2$. The exact phases 
should not matter for computing most quantities and we adopt the values
used by 
\cite{2003ApJ...591.1129A}.

Mode displacements are computed from the equation
\footnote{The sign of the last term here is opposite to Eq. (29) in
\cite{2003ApJ...591.1129A} because of the different sign convention for frequency
used here.}
\be
\left(1-{1\over\mu^2}\right)\xivec
=\grad\Psi-{\zhat\zhat\dotprod\grad\Psi\over\mu^2}+{i\zhat\crossprod
\grad\Psi\over\mu}
\ee
up to an overall normalization factor. With the approximation that
the density scale height is large, we do not include derivatives
of $\rho$ in computing the displacements; thus we write
\ba
\sqrt{\rho}\left(1-{1\over\mu^2}\right)\xivec
&\approx&\exp(+i\omega t)\Biggl\{
\grad\left[P_{nm}(x_1)P_{nm}(x_2)\exp(im\phi)\right]
\nonumber\\& &
-{\zhat\zhat\dotprod\grad\left[P_{nm}(x_1)P_{nm}(x_2)\exp(im\phi)\right]
\over\mu^2}\nonumber\\& &
+{i\zhat\over\mu}\crossprod
\grad\left[P_{nm}(x_1)P_{nm}(x_2)\exp(im\phi)\right]\Biggr\}~.
\label{xiwkbone}
\ea
For evaluating the derivatives, we use
\ba
\grad P_{nm}(x_i)&=&{dP_{nm}\over dx_i}\grad x_i
\nonumber\\
\grad x_i&=&-\sin\theta_i\grad\theta_i=-\sqrt{1-x_i^2}\grad\theta_i
\ea
where the $\grad\theta_i$ were computed in Eqs. (\ref{thetaderivscyl}).
If we further invoke the large $p$ approximation to the
associated Legendre polynomials, then we ignore the variation of the
$\sin\theta_i$ factors in computing derivatives; in this approximation
\ba
\sqrt{\rho\sin\theta_1\sin\theta_2}\left(1-{1\over\mu^2}\right)\xivec
&\approx&\exp(+i\omega t)\Biggl\{
\grad\left[\cos(p\theta_1+\alpha_1)\cos(p\theta_2+\alpha_2)\exp(im\phi)\right]
\nonumber\\& &
-{\zhat\zhat\dotprod\grad\left[\cos(p\theta_1+\alpha_1)
\cos(p\theta_2+\alpha_2)\exp(im\phi)\right]\over\mu^2}\nonumber\\& &
+{i\zhat\over\mu}\crossprod
\grad\left[\cos(p\theta_1+\alpha_1)\cos(p\theta_2+\alpha_2)
\exp(im\phi)\right]\Biggr\}~.
\label{xiwkbtwo}
\ea
In Eq. (\ref{xiwkbtwo}) gradients are computed via
\be
\grad_a[\cos(p\theta_i+\alpha_i)]=-p\grad_a\theta_i\sin(p\theta_1+\alpha_i)
\ee
where $\grad\theta_i$ are computed from Eqs. (\ref{thetaderivscyl}).
%
The components of the displacement are
\ba
\left(1-{1\over\mu^2}\right)\xivec_{\rperp}&=&{\partial\Psi\over\partial
\rperp}+{im\Psi\over\mu\rperp}\approx\frac{\partial\Psi}{\partial\varpi}
\nonumber\\
\xivec_z&=&
{\partial\Psi\over\partial z}
\nonumber\\
\left(1-{1\over\mu^2}\right)\xivec_\phi&=&{im\Psi\over\rperp}
+{i\over\mu}{\partial\Psi\over\partial\rperp}\approx\frac{i\xi_\varpi}{\mu}~,
\label{displacements}
\ea
where the approximations are valid within the WKB limit. The necessary
derivatives are
\ba
\frac{\partial\Psi}{\partial\rperp}&=&-\frac{p\sqrt{1-\mu^2}
e^{i(m\phi+\omega t)}}{2\sqrt{\rho\sin\theta_1\sin\theta_2}}
\left(\frac{\sin\eta_+}{\sin\ttheta_+}-\frac{\sin\eta_-}{\sin\ttheta_-}\right)
\nonumber\\
\frac{\partial\Psi}{\partial z}&=&\frac{p\vert\mu\vert e^{i(m\phi+\omega t)}}{2\sqrt{\rho\sin\theta_1\sin\theta_2}}
\left(\frac{\sin\eta_+}{\sin\ttheta_+}+\frac{\sin\eta_-}{\sin\ttheta_-}\right)
\label{xicomps}
\ea
where $\eta_\pm=p(\theta_2\pm\theta_1)+\alpha_2\pm\alpha_1$ and
$\ttheta_\pm=\theta_2\pm\theta_1$.

\subsection{Normalization Integral}

Define 
\be
N\equiv\int d^3x \rho\vert\xivec\vert^2~;
\ee
using Eqs. (\ref{displacements}) and (\ref{xicomps}) as well as 
Eq. (\ref{useful}) we get
\be
N=\frac{\pi\vert\mu\vert p^2}{(1-\mu^2)^2}\int d\theta_1 d\theta_2
\left(\frac{\sin^2\eta_+\sin\ttheta_-}{\sin\ttheta_+}+\frac{\sin^2\eta_-
\sin\ttheta_+}{\sin\ttheta_-}-2\mu^2\sin\eta_+\sin\eta_-\right)~.
\ee
We replace the rapidly oscillating terms $\sin^2\eta_\pm\to\onehalf$ and
$\sin\eta_+\sin\eta_-\to 0$. 
Judiciously 
substitute $\sin\ttheta_\pm=\sin(\ttheta_\mp\pm 2\theta_1)
=\sin\ttheta_\mp\cos 2\theta_1\pm\cos\ttheta_\mp\sin 2\theta_1$, with
which the integral becomes
\be
N={\pi p^2\muvert\over 2(1-\mu^2)^2}\int_0^{\thetamu}d\theta_1
\int_{\thetamu}^{\pi-\thetamu}d\theta_2\left[2\cos 2\theta_1
+\sin 2\theta_1\left({\cos\ttheta_-\over\sin\ttheta_-}
-{\cos\ttheta_+\over\sin\ttheta_+}\right)\right]~.
\ee
The remaining integrals may all be done analytically; the result is
\be
N=\frac{\pi^2p^2\mu^2}{(1-\mu^2)^{3/2}}~.
\label{modenorm}
\ee
See 
\cite{2003ApJ...591.1129A}, Eq. (47); the exact result for Bryan modes is
in 
\cite{2004PhRvD..70l4017B}, \S II.D.

\subsection{Damping in a Viscous Bondary Layer}
\label{blAppendix}

For evaluating damping via boundary layer viscosity, we will need to
compute the surface integral of $\rho\vert\xivec\vert^2$. We use
Eq. (\ref{thetapm}) to write
$\ttheta_\pm=\theta_2\pm\theta_1=\cos^{-1}(r\cos\theta_\pm)$
and therefore
\ba
{\rho\vert\xivec\vert^2}&\approx&\frac
{p^2\mu^2}{2(1-\mu^2)\sin\theta_1\sin\theta_2}
\left\{{\sin^2[p\cos^{-1}(r\cos\thetap)-p\pi]\over
1-r^2+r^2\sin^2\thetap}
+{\sin^2[p\cos^{-1}(r\cos\thetam)]\over
1-r^2+r^2\sin^2\thetam}
\right\}
\nonumber\\& &
-{\sigma\mu^4\sin[p\cos^{-1}(r\cos\thetap)-p\pi]
\sin[p\cos^{-1}(r\cos\thetam)]\over(1-\mu^2)\sin\theta_1
\sin\theta_2
\sin[\cos^{-1}(r\cos\thetap)]\sin[\cos^{-1}(\cos\thetam)]}~.
\ea
where $\sigma=+1$ for even parity and $\sigma=-1$ for odd parity, and
in the first two terms we used $\sin^2[\cos^{-1}(r\cos\theta_\pm)]
=1-r^2\cos^2\theta_\pm=1-r^2+r^2\sin^2\theta_\pm$. Using
Eq. (\ref{setup}) to eliminate $\sin\theta_1\sin\theta_2$ 
we get
\ba
\frac{r^2}{2N}\int d\Omega\rho\vert\xivec^2\vert&\approx&
\frac{r}{2\pi}\int_0^\pi d\theta
\left\{\frac{\sin^2\{p[\cos^{-1}(r\cos\thetap)-\pi\}}
{1-r^2+r^2\sin^2\thetap}
+\frac{\sin^2[p\cos^{-1}(r\cos\thetam)]}
{1-r^2+r^2\sin^2\thetam}
\right\}
\nonumber\\& &
-\frac{\sigma r}{\pi}
\int_0^\pi\frac{d\theta\sin\{p[\cos^{-}(r\cos\thetap)-\pi\}
\sin[p\cos^{-1}(r\cos\thetam)]}
{\sqrt{(1-r^2+r^2\sin^2\thetap)(1-r^2+r^2\sin^2\thetam)}}
\label{xisqsurfintegral}
\ea
where $\sigma=+1$ for even parity and $\sigma=-1$ for odd parity.
In the notation of Eq. (\ref{crampitval}), $R(dE_D/dr)_b/E_D$ is
{\sl twice} Eq. (\ref{xisqsurfintegral}).
In Eq. (\ref{xisqsurfintegral}), the integrands in $\{\cdots\}$ have large
values near $\theta_+=\pi$ and $\theta_-=0$, respectively.
Writing $\theta_+=\pi+\delta$ and $\theta_-=\delta$, we get
$1-r^2+r^2\sin^2\theta_\pm\approx 1-r^2+r^2\delta^2$, and
$\cos^{-1}(r\cos\theta_+)=\pi-\sqrt{1-r^2+r^2\delta^2}$ and
$\cos^{-1}(r\cos\theta_-)=\sqrt{1-r^2+r^2\delta^2}$, respectively, near those points;
in both cases, then
\be
{\sin^2[p\cos^{-1}(r\cos\theta)]\over 1-r^2+r^2\sin^2\theta}
\approx{\sin^2[p\sqrt{1-r^2+r^2\delta^2}]\over1-r^2+r^2\delta^2}
={\sin^2[p\sqrt{1-r^2}(1+u^2)^{1/2}]\over(1-r^2)(1+u^2)}\,,
\ee
where $u^2=r^2\delta^2/(1-r^2)$.
If $1-r^2\ll 1$ we can approximate the integrals by
\be
{1\over r\sqrt{1-r^2}}\int_{-\infty}^{+\infty}
{du\sin^2[p\sqrt{1-r^2}(1+u^2)^{1/2}]\over1+u^2}
\equiv {\pi K(p\sqrt{1-r^2})\over r\sqrt{1-r^2}}~;
\ee
\be
K(z)\approx\left\{\begin{array}{ll}
     z &\mbox{if $z\ll 1$}\\
     \onehalf &\mbox{if $z\gg 1$}
         \end{array}
         \right.
\ee
Since the contribution to the integral from the cross term
is smaller, we find that
\be
\frac{r^2}{2N}\int d\Omega\rho\vert\xivec^2\vert\approx
{K(p\sqrt{1-r^2})\over\sqrt{1-r^2}}
\label{xisqsurfans}
\ee
for $1-r^2\ll 1$. The damping rate from a viscous shearing boundary layer
is therefore proportional to $R(dE_D/dr)_b/2E_D\simeq K(p\sqrt{1-r^2})/\sqrt{1-r^2}$; at large
$p$, the damping rate is proportional to $1/2\sqrt{1-r^2}$, which is
independent of $p$. 

\subsection{Shear Viscosity Damping within $r=r_b$}
\label{ShearViscosityrb}
For computing shear viscosity damping, we need the square of the shear tensor:
\be
\sigma_{ab}=\frac{\partial\xi_a}{\partial x_b}
+\frac{\partial\xi_b}{\partial x_b}
-\frac{2\delta_{ab}\div\xivec}{3}
\approx\frac{\partial\xi_a}{\partial x_b}+\frac{\partial\xi_b}{\partial x_b}
\equiv S_{ab}+S_{ba}
\ee
where the approximation $\div\xivec=0$ holds in WKB. Using Eq. (\ref{displacements})
we can compute the shear tensor components; then
\be
\sigma^2\equiv\sum_{ab}\sigma_{ab}\sigma_{ab}
=\left[\frac{8\mu^4+2\mu^2}{(1-\mu^2)^2}\right]\left(\frac{\partial^2\Psi}
{\partial\varpi^2}\right)^2+\frac{2(1-3\mu^2+4\mu^4)}{(1-\mu^2)^2}
\left(\frac{\partial^2\Psi}{\partial\varpi\partial z}\right)^2~.
\label{sigsq}
\ee
In the WKB limit
\ba
\frac{\partial^2\Psi}{\partial u\partial v}=-\frac{p^2}{\sqrt{\rho\sin\theta_1
\sin\theta_2}}\Biggl[\cos(p\theta_1+\alpha_1)\cos(p\theta_2+\alpha_2)
\left(\frac{\partial\theta_1}{\partial u}\frac{\partial\theta_1}{\partial v}
+\frac{\partial\theta_2}{\partial u}\frac{\partial\theta_2}{\partial v}\right)
\nonumber\\
-\sin(p\theta_1+\alpha_1)\sin(p\theta_2+\alpha_2)\left(\frac{\partial\theta_1}
{\partial u}\frac{\partial\theta_2}{\partial v}+\frac{\partial\theta_1}{\partial v}
\frac{\partial\theta_2}{\partial u}\right)\Biggr]~;
\ea
using Eqs. (\ref{thetaderivscyl})
and 
$\sin^2(\theta_1\pm\theta_2)=1-r^2\cos^2\theta_\pm$
we get 
\ba
\sigma^2=\frac{p^4\mu^2}{2\rho r\sin\theta(1-\mu^2)^{3/2}}\Biggl\{\frac{\cos^2[p(\theta_1+\theta_2)+\alpha_1+\alpha_2]}
{
(1-r^2\cos^2\theta_+)^2}+\frac{\cos^2[p(\theta_1-\theta_2)+\alpha_1-\alpha_2]}{
(1-r^2\cos^2\theta_-)^2}
\nonumber\\
+\frac{(4\mu^2(3-4\mu^2)\cos^2[p(\theta_1+\theta_2)+\alpha_1+\alpha_2]
\cos^2[p(\theta_1-\theta_2)+\alpha_1-\alpha_2]}{
(1-r^2\cos^2\theta_+)(1-r^2\cos^2\theta_-)}\Biggr\}
~;
\label{sigsqfinal}
\ea
the terms on the second line of Eq. (\ref{sigsqfinal}) oscillate rapidly and will be dropped in our detailed
calculation of the damping rate.

We will assume that the main contribution to the damping rate is from the core of the neutron star, $r\leq
r_b<1$; this means that we will never encounter the {\sl exactly two} points on the surface where Eq. (\ref{sigsqfinal})
is singular. Fig. 6 in 
\cite{2008PhRvD..78f3006S}
suggests that the shear viscosity grows perhaps linearly in the core
of a neutron star, so we let $\eta=\nucore\rho$ in the core, where $\nucore$ is independent of density; this roughly cancels the $1/\rho$ factor in Eq. (\ref{sigsqfinal}) from the WKB form of the
modes. For large values of $p$, we approximate the first two terms in Eq. (\ref{sigsqfinal}) by replacing
$\cos^2[p(\theta_1\pm\theta_2)+\alpha_1\pm\alpha_2]\to\onehalf$, and drop the cross term entirely. Then
\ba
\int d^3r\rho\sigma^2&\approx&\frac{\pi p^4\mu^2}{2(1-\mu^2)^{3/2}}\int_0^\pi d\theta\int_0^{r_b} dr\,r
\left[\frac{1}{(1-r^2\cos^2\theta_+)^2}+\frac{1}{(1-r^2\cos^2\theta_-)^2}\right]
\nonumber\\
&=&\frac{\pi p^4r_b^2\mu^2}{4(1-\mu^2)^{3/2}}\sum_{s=\pm}
\int_0^{2\pi}\frac{d\theta}{1-r_b\cos\theta_s}
\ea
The integrals involved are all $2\pi/\sqrt{1-r_b^2}$, so the final result is
\be
\int d^3r\rho\sigma^2=\frac{\pi^2 p^4r_b^2\mu^2}{(1-\mu^2)^{3/2}\sqrt{1-r_b^2}}
\label{sigint}
\ee
Dividing by $N$
and multiplying by $\nucore$ gives the damping rate
\be
\gamma_{core}=\frac{\nucore p^2(r_b/R)^2}{R^2\sqrt{1-(r_b/R)^2}}
\label{gammacore}
\ee
where we have restored dimensional units. Note that Eq. (\ref{gammacore})
would diverge as $r_b\to R$. That case requires a more careful treatement.

\cite{2004PhRvD..70l4017B} included the entire star in the calculation of shear
damping: Eq. (31) in that paper is an accurate analytic fit, which we
reproduce here: for kinematic viscosity $\eta$,
\be
\frac{\gamma R^2}{\eta}=\frac{2n+1}{3}\left[(n+3)(n-2)-\frac{m(m-2\mu)}{1-\mu^2}
\right]
\ee
which is approximately $\gamma/\eta=2n^3/3$ for $n\gg\vert m\vert$, which
is typical of couplings of large $n$ modes to the R mode. We have also done
a WKB calculation that gives $\gamma R^2/\eta\approx 2p^3/3$. That calculation
is rather complicated because the result is dominated by contributions
from near the special points on the surface where $\cos\theta_1=\pm\cos\theta_2
=\muvert$. The procedure is to return to the displacement field $\xivec$,
introduce approximations valid near the special points, and then compute
the shear tensor by direct differentiation. This last step deviates from
the strict WKB approximation in that if $\kvec_i=\grad\theta_i$ it includes
terms arising from $\grad\kvec_i$ that would be discarded ordinarily. The
expression that results can then be integrated analytically, and the
result is what we quoted above.

To get the expression for shear viscosity damping in the main text, we
divide the star into a core out to $r_b$ and crust outside $r_b$, with
separate viscosities $\nucore$ and $\nucrust$, respectively.

\def\ddx{\frac{\partial}{\partial x}}
\def\dxidx{\frac{\partial\xi}{\partial x}}
\def\ctS{c_{t,S}^2}
\def\ddu{\frac{\partial}{\partial u}}
\def\dxidu{\frac{\partial\xi}{\partial u}}
\def\nuunit{\nu_{500}}
\def\cms{\,{\rm cm\,s^{-1}}\,}
\def\etahat{{\hat\eta}}

\section{Toy Model for Displacement Evolution as Shear Modulus Rises}
\label{toyshear}

We consider the transition region of thickness $\Delta r$ within which the 
shear modulus $\mu(x)$ rises from zero to its value in the crystalline crust. We
ignore density variation, and consider planar displacement fields only
with $\div\xivec=0$. We orient the radial direction along $x$ and define
$c_t^2(x)=\mu(x)/\rho$.

We assume that displacements are proportional to functions of $x$ times
$\exp[i(k_yy+k_zz)-i\omega t]$; the divergenceless condition implies
that $\xi_y$ and $\xi_z$ are both ${\cal O}(\vert\partial\xi_x/\partial x\vert)$
and hence much larger than $\xi_x$. Then if we systematically ignore $k_{y,z}$
compared with $\partial/\partial x$ in this region both $\xi_y$ and $\xi_z$
obey the approximate linear differential equation
\be
-\omega^2\xi=\ddx\left(c_t^2\dxidx\right)~.
\label{xieqn}
\ee
This equation describes the evolution of the jumps in these displacement components.
Let $c_t^2=\ctS f(u)$, where within the layer $x=x_{inner}+u\Delta r$ and in 
the solid $c_t^2=\ctS$. The function $f(u)$ may be determined from microphysics.
Written in terms of $u$ Eq. (\ref{xieqn}) is
\be
0=\ddu\left[f(u)\dxidu\right]+\frac{\omega^2(\Delta r)^2\xi}{\ctS}
\equiv\ddu\left[f(u)\dxidu\right]+q^2\xi~.
\label{xieqnnondim}
\ee
Provided that both $f(u)$ and $\xi(u)$ are monotonic, we can regard $\xi$
as a function of $f$.

Realistically, we would solve Eq. (\ref{xieqnnondim}) for a specified $f(u)$.
To get a rough idea of what a solution might look like, we pursue an illustrative
toy calculation: left $\xi=f^p$, where $p$ is some powerlaw index, to get
\be
0=\frac{d^2f^{1+p}}{du^2}+q^2(1+p)f^p~;
\ee
rescale so that $g=Af^{1+p}$ to get
\be
0=\frac{d^2g}{du^2}+g^{\frac{p}{1+p}}
\label{geqn}
\ee
where we have chosen the scaling constant so that $q^2(1+p)A^{\frac{1}{1+p}}=1$.
We solve Eq. (\ref{geqn}) with $g=0$ at $u=0$ but $(dg/du)_0\neq 0$; we impose
the condition that $(dg/du)_1=0$ at $u=1$, which follows since $(df/du)_1=0$ for $p>0$.
Consequently $(dg/du)_0$ is an eigenvalue. Since we require that $f(1)=1$, it follows
that $g(1)=A$, so we have $q^2(1+p)[g(1)]^{\frac{1}{1+p}}=1$, which determines $q^2$.
Thus, it should be clear that this choice of $\xi(f)$ is hardly general, and would
only hold for a very specific $f(u)$ and $q^2$.

Note that Eq. (\ref{geqn}) can be integrated once to yield
\be
\frac{dg}{du}=\left(\frac{dg}{du}\right)_0\sqrt{1-\left(\frac{g}{g_0}\right)^{\frac{1+2p}{1+p}}}
\label{gintegrated}
\ee
where $g_0^{\frac{1+2p}{1+p}}\equiv (1+2p)(dg/du)_0^2/2(1+p)$.
Eq. (\ref{gintegrated}) can be solved via quadruture. An acceptable solution has $g(1)=g_0$. 

The viscous dissipation rate within the layer is
\be
\dot E=4\pi\rho_br_b^2\omega^2\int dx\,\xi\ddx\left(\eta\dxidx\right)=\frac{4\pi\rho_b r_b^2\omega^2}{\Delta r}
\left[\xi\eta(u)\dxidu\biggr\vert_0^1-\int_0^1du\eta(u)\left(\dxidu\right)^2\right]~.
\ee
Let the kinemaic viscosity be $\eta(u)=\etacore\etahat(u)$, where $\etahat(0)=1$ and $\etahat(1)=\etacrust/\etacore
\ll 1$; we assume $\etahat(u)\leq 1$ to get an upper bound on $\dot E$  Since
$\omega\xi=\Delta v[g/g(1)]^{\frac{p}{1+p}}$ in our toy model
\be
\dot E=-\frac{4\pi p^2\etacore\rho_br_b^2(\Delta v)^2}{\Delta r(1+p)^2[g(1)]^{\frac{2p}{1+p}}}
\left\{\left[\frac{(1+p)g^{\frac{p-1}{1+p}}}{p}\frac{dg}{du}\right]_0+\int_0^1\frac{du\etahat(u)}{[g(u)]^{\frac{2}{1+p}}}
\left(\frac{dg(u)}{du}\right)^2\right\}~.
\label{transdiss}
\ee
Since $g(u)\sim u$ at $u\ll 1$ the integral diverges for $p\leq 1$

\begin{figure}
\begin{center}
\epsfxsize = 200pt
\epsfbox{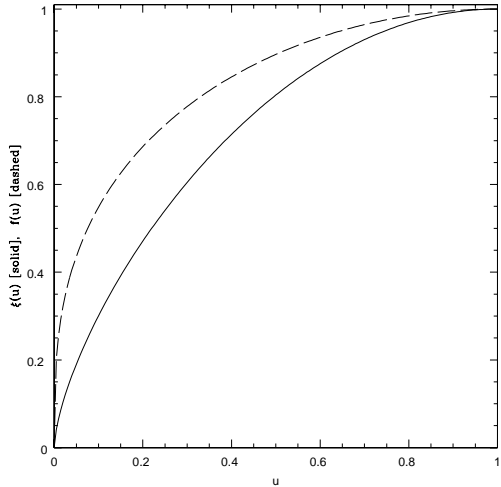}
\caption{$\xi(u)$ [solid] and $f(u)$ [dashed] for the toy model with $p=2$. The solution
shows that the displacement field changes smoothly within the transition zone from crust
to core, with a characteristic length scale $\sim\Delta r$, the thickness of the zone.}
\label{fig:toy}
\end{center}
\end{figure}

We have solved Eq. (\ref{geqn}) for $p=2$; the eigenvalue is
$(dg/du)_0\approx 0.135164405635$, and $g(1)\approx 0.0811944$;
consequently, $3\omega^2(\Delta r)^2[g(1)]^{1/3}/\ctS=1$, or
$c_{t,S}=1.14\omega\Delta r=3.5\times 10^7\cms\nuunit(\omega/\Omega)(\Delta r/100\,{\rm m})$,
which is a plausible value but cannot be right for all modes, each of which has its
own value of $\omega$. For a given $f(u)$ the function $\xi(f)$ must differ among modes
and generally the problem does not scale as it does when $\xi(f)=f^p$.
Nevertheless, this toy model illustrates the salient features of how a transition might occur. 
The solution is shown in Fig. \ref{fig:toy}. The dissipation rate in this
model is $\dot E\leq 4\pi\etacore\rho_br_b^2(\Delta v)^2/\Delta r\times 1.99$.

\acknowledgments We thank J. Brink, J. Cordes, D. Charkrabarty and P. Shternin for helpful correspondence. 
IW is grateful for past discussions with P. Arras and S. Teukolsky on some of the approximations
required to derive the WKB results in Appendix A.
RB is grateful to L. S. Finn, B. Owen and P. Jetzer for useful discussions and support. 
This research was supported in part by NASA ATP grant NNX13AH42G
to Cornell University
and by NSF grant PHY11-25915 to the Kavli Institute for Theoretical Physics at
University of California, Santa Barbara. RB aknowledges current support from the Dr. Tomalla Foundation
and the Swiss National Science Foundation. She was previously supported by NSF PHY 09-69857 awarded to the 
Pennsylvania State University.


\begin{thebibliography}{67}
\expandafter\ifx\csname natexlab\endcsname\relax\def\natexlab#1{#1}\fi

\bibitem[{{Alpar} {et~al.}(1982){Alpar}, {Cheng}, {Ruderman}, \&
  {Shaham}}]{1982Natur.300..728A}
{Alpar}, M.~A., {Cheng}, A.~F., {Ruderman}, M.~A., \& {Shaham}, J. 1982,
  Nature, 300, 728

\bibitem[{{Andersson}(1998)}]{1998ApJ...502..708A}
{Andersson}, N. 1998, \apj, 502, 708, arXiv:gr-qc/9706075

\bibitem[{{Andersson} {et~al.}(2005){Andersson}, {Comer}, \&
  {Glampedakis}}]{2005NuPhA.763..212A}
{Andersson}, N., {Comer}, G.~L., \& {Glampedakis}, K. 2005, Nuclear Physics A,
  763, 212, arXiv:astro-ph/0411748

\bibitem[{{Andersson} {et~al.}(1999){Andersson}, {Kokkotas}, \&
  {Schutz}}]{1999ApJ...510..846A}
{Andersson}, N., {Kokkotas}, K., \& {Schutz}, B.~F. 1999, \apj, 510, 846,
  arXiv:astro-ph/9805225

\bibitem[{{Arras} {et~al.}(2003){Arras}, {Flanagan}, {Morsink}, {Schenk},
  {Teukolsky}, \& {Wasserman}}]{2003ApJ...591.1129A}
{Arras}, P., {Flanagan}, E.~E., {Morsink}, S.~M., {Schenk}, A.~K., {Teukolsky},
  S.~A., \& {Wasserman}, I. 2003, \apj, 591, 1129, arXiv:astro-ph/0202345

\bibitem[{{Arzoumanian} {et~al.}(1999){Arzoumanian}, {Cordes}, \&
  {Wasserman}}]{1999ApJ...520..696A}
{Arzoumanian}, Z., {Cordes}, J.~M., \& {Wasserman}, I. 1999, \apj, 520, 696,
  arXiv:astro-ph/9811323

\bibitem[{{Bildsten}(1998)}]{1998ApJ...501L..89B}
{Bildsten}, L. 1998, \apjl, 501, L89+, arXiv:astro-ph/9804325

\bibitem[{{Bildsten} {et~al.}(1997){Bildsten}, {Chakrabarty}, {Chiu}, {Finger},
  {Koh}, {Nelson}, {Prince}, {Rubin}, {Scott}, {Stollberg}, {Vaughan},
  {Wilson}, \& {Wilson}}]{1997ApJS..113..367B}
{Bildsten}, L. {et~al.} 1997, \apjs, 113, 367, arXiv:astro-ph/9707125

\bibitem[{{Bildsten} \& {Ushomirsky}(2000)}]{2000ApJ...529L..33B}
{Bildsten}, L., \& {Ushomirsky}, G. 2000, \apjl, 529, L33,
  arXiv:astro-ph/9911155

\bibitem[{{Bondarescu} {et~al.}(2007){Bondarescu}, {Teukolsky}, \&
  {Wasserman}}]{2007PhRvD..76f4019B}
{Bondarescu}, R., {Teukolsky}, S.~A., \& {Wasserman}, I. 2007, \prd, 76,
  064019, 0704.0799

\bibitem[{{Bondarescu} {et~al.}(2009){Bondarescu}, {Teukolsky}, \&
  {Wasserman}}]{2009PhRvD..79j4003B}
------. 2009, \prd, 79, 104003, 0809.3448

\bibitem[{{Boutloukos} \& {Nollert}(2007)}]{2007PhRvD..75d3007B}
{Boutloukos}, S., \& {Nollert}, H.-P. 2007, \prd, 75, 043007,
  arXiv:gr-qc/0605044

\bibitem[{{Brink}(2005)}]{2005PhDT........26B}
{Brink}, J. 2005, PhD thesis, Cornell University, United States -- New York

\bibitem[{{Brink} {et~al.}(2004){Brink}, {Teukolsky}, \&
  {Wasserman}}]{2004PhRvD..70l4017B}
{Brink}, J., {Teukolsky}, S.~A., \& {Wasserman}, I. 2004, \prd, 70, 124017,
  arXiv:gr-qc/0409048

\bibitem[{{Brink} {et~al.}(2005){Brink}, {Teukolsky}, \&
  {Wasserman}}]{2005PhRvD..71f4029B}
------. 2005, \prd, 71, 064029, arXiv:gr-qc/0410072

\bibitem[{{Brown}(2000)}]{2000ApJ...531..988B}
{Brown}, E.~F. 2000, \apj, 531, 988, arXiv:astro-ph/9910215

\bibitem[{{Chakrabarty}(2005)}]{2005AIPC..797...71C}
{Chakrabarty}, D. 2005, in American Institute of Physics Conference Series,
  Vol. 797, Interacting Binaries: Accretion, Evolution, and Outcomes, ed.
  {L.~Burderi, L.~A.~Antonelli, F.~D'Antona, T.~di Salvo, G.~L.~Israel,
  L.~Piersanti, A.~Tornamb{\`e}, \& O.~Straniero}, 71--80

\bibitem[{{Chakrabarty}(2008)}]{2008AIPC.1068...67C}
{Chakrabarty}, D. 2008, in American Institute of Physics Conference Series,
  Vol. 1068, American Institute of Physics Conference Series, ed. {R.~Wijnands,
  D.~Altamirano, P.~Soleri, N.~Degenaar, N.~Rea, P.~Casella, A.~Patruno, \&
  M.~Linares}, 67--74

\bibitem[{{Chakrabarty}(2012)}]{2012cosp...39..295C}
{Chakrabarty}, D. 2012, in COSPAR Meeting, Vol.~39, 39th COSPAR Scientific
  Assembly, 295

\bibitem[{{Chandrasekhar}(1970)}]{1970PhRvL..24..611C}
{Chandrasekhar}, S. 1970, Physical Review Letters, 24, 611

\bibitem[{{Cook} {et~al.}(1994){Cook}, {Shapiro}, \&
  {Teukolsky}}]{1994ApJ...423L.117C}
{Cook}, G.~B., {Shapiro}, S.~L., \& {Teukolsky}, S.~A. 1994, \apjl, 423, L117+

\bibitem[{{Cutler} {et~al.}(1990){Cutler}, {Lindblom}, \&
  {Splinter}}]{1990ApJ...363..603C}
{Cutler}, C., {Lindblom}, L., \& {Splinter}, R.~J. 1990, \apj, 363, 603

\bibitem[{{Flowers} {et~al.}(1976){Flowers}, {Ruderman}, \&
  {Sutherland}}]{1976ApJ...205..541F}
{Flowers}, E., {Ruderman}, M., \& {Sutherland}, P. 1976, \apj, 205, 541

\bibitem[{{Friedman} \& {Morsink}(1998)}]{1998ApJ...502..714F}
{Friedman}, J.~L., \& {Morsink}, S.~M. 1998, \apj, 502, 714,
  arXiv:gr-qc/9706073

\bibitem[{{Friedman} \& {Schutz}(1978)}]{1978ApJ...222..281F}
{Friedman}, J.~L., \& {Schutz}, B.~F. 1978, \apj, 222, 281

\bibitem[{{Glampedakis} \& {Andersson}(2006)}]{2006MNRAS.371.1311G}
{Glampedakis}, K., \& {Andersson}, N. 2006, \mnras, 371, 1311,
  arXiv:astro-ph/0607105

\bibitem[{{Gusakov} {et~al.}(2004){Gusakov}, {Kaminker}, {Yakovlev}, \&
  {Gnedin}}]{2004A&A...423.1063G}
{Gusakov}, M.~E., {Kaminker}, A.~D., {Yakovlev}, D.~G., \& {Gnedin}, O.~Y.
  2004, \aap, 423, 1063, arXiv:astro-ph/0404002

\bibitem[{{Haskell} {et~al.}(2009){Haskell}, {Andersson}, \&
  {Passamonti}}]{2009MNRAS.397.1464H}
{Haskell}, B., {Andersson}, N., \& {Passamonti}, A. 2009, \mnras, 397, 1464,
  0902.1149

\bibitem[{{Haskell} {et~al.}(2013){Haskell}, {Glampedakis}, \& {Andersson}}]{Haskell13}
{Haskell}, B., {Glampedakis}, K., \& {Andersson}, N. 2013, 1307.0985.
 
\bibitem[{{Hessels} {et~al.}(2006){Hessels}, {Ransom}, {Stairs}, {Freire},
  {Kaspi}, \& {Camilo}}]{2006Sci...311.1901H}
{Hessels}, J.~W.~T., {Ransom}, S.~M., {Stairs}, I.~H., {Freire}, P.~C.~C.,
  {Kaspi}, V.~M., \& {Camilo}, F. 2006, Science, 311, 1901,
  arXiv:astro-ph/0601337
  
  \bibitem[{{Heyl} (2002){Heyl}}]{Heyl}
{Heyl}, J. 2002, \apjl, 574, L57.  arXiv:astro-ph/0206174

  
\bibitem[Ivanov \& Papaloizou(2010)]{2010MNRAS.407.1609I} Ivanov, P.~B., \& Papaloizou, J.~C.~B.\ 2010, \mnras, 407, 1609 

\bibitem[{{Kinney} \& {Mendell}(2003)}]{2003PhRvD..67b4032K}
{Kinney}, J.~B., \& {Mendell}, G. 2003, \prd, 67, 024032, arXiv:gr-qc/0206001

\bibitem[{{Kojima}(1998)}]{1998MNRAS.293...49K}
{Kojima}, Y. 1998, \mnras, 293, 49, arXiv:gr-qc/9709003

\bibitem[{{Kojima} \& {Hosonuma}(1999)}]{1999ApJ...520..788K}
{Kojima}, Y., \& {Hosonuma}, M. 1999, \apj, 520, 788, arXiv:astro-ph/9903055

\bibitem[{{Kolomeitsev} \& {Voskresensky}(2008)}]{2008PhRvC..77f5808K}
{Kolomeitsev}, E.~E., \& {Voskresensky}, D.~N. 2008, \prc, 77, 065808,
  0802.1404

\bibitem[{{Lee} \& {Strohmayer}(1996)}]{1996A&A...311..155L}
{Lee}, U., \& {Strohmayer}, T.~E. 1996, \aap, 311, 155

\bibitem[{{Levin} \& {Ushomirsky}(2001)}]{2001MNRAS.324..917L}
{Levin}, Y., \& {Ushomirsky}, G. 2001, \mnras, 324, 917, arXiv:astro-ph/0006028

\bibitem[{{Lindblom} \& {Owen}(2002)}]{2002PhRvD..65f3006L}
{Lindblom}, L., \& {Owen}, B.~J. 2002, \prd, 65, 063006, arXiv:astro-ph/0110558

\bibitem[{{Lindblom} {et~al.}(1998){Lindblom}, {Owen}, \&
  {Morsink}}]{1998PhRvL..80.4843L}
{Lindblom}, L., {Owen}, B.~J., \& {Morsink}, S.~M. 1998, Physical Review
  Letters, 80, 4843, arXiv:gr-qc/9803053

\bibitem[{Lockitch {et~al.}(2000)Lockitch, Andersson, \&
  Friedman}]{PhysRevD.63.024019}
Lockitch, K.~H., Andersson, N., \& Friedman, J.~L. 2000, Phys. Rev. D, 63,
  024019

\bibitem[{{Lockitch} {et~al.}(2001){Lockitch}, {Andersson}, \&
  {Friedman}}]{2001PhRvD..63b4019L}
{Lockitch}, K.~H., {Andersson}, N., \& {Friedman}, J.~L. 2001, \prd, 63,
  024019, arXiv:gr-qc/0008019

\bibitem[{{Lockitch} {et~al.}(2004){Lockitch}, {Andersson}, \&
  {Watts}}]{2004CQGra..21.4661L}
{Lockitch}, K.~H., {Andersson}, N., \& {Watts}, A.~L. 2004, Classical and
  Quantum Gravity, 21, 4661, arXiv:gr-qc/0106088

\bibitem[{{Lockitch} \& {Friedman}(1999)}]{1999ApJ...521..764L}
{Lockitch}, K.~H., \& {Friedman}, J.~L. 1999, \apj, 521, 764,
  arXiv:gr-qc/9812019

\bibitem[{Lockitch {et~al.}(2003)Lockitch, Friedman, \&
  Andersson}]{PhysRevD.68.124010}
Lockitch, K.~H., Friedman, J.~L., \& Andersson, N. 2003, Phys. Rev. D, 68,
  124010

\bibitem[{{Manchester} {et~al.}(2005){Manchester}, {Hobbs}, {Teoh}, \&
  {Hobbs}}]{2005AJ....129.1993M}
{Manchester}, R.~N., {Hobbs}, G.~B., {Teoh}, A., \& {Hobbs}, M. 2005, \aj, 129,
  1993

\bibitem[{{Mendell}(2001)}]{2001PhRvD..64d4009M}
{Mendell}, G. 2001, \prd, 64, 044009, arXiv:gr-qc/0102042

\bibitem[{{Nayyar} \& {Owen}(2006)}]{2006PhRvD..73h4001N}
{Nayyar}, M., \& {Owen}, B.~J. 2006, \prd, 73, 084001, arXiv:astro-ph/0512041

\bibitem[{{Page} {et~al.}(2004){Page}, {Lattimer}, {Prakash}, \&
  {Steiner}}]{2004ApJS..155..623P}
{Page}, D., {Lattimer}, J.~M., {Prakash}, M., \& {Steiner}, A.~W. 2004, \apjs,
  155, 623, arXiv:astro-ph/0403657

\bibitem[{{Page} {et~al.}(2009){Page}, {Lattimer}, {Prakash}, \&
  {Steiner}}]{2009arXiv0906.1621P}
------. 2009, ArXiv e-prints, 0906.1621

\bibitem[{{Page} {et~al.}(2011){Page}, {Prakash}, {Lattimer}, \&
  {Steiner}}]{2011PhRvL.106h1101P}
{Page}, D., {Prakash}, M., {Lattimer}, J.~M., \& {Steiner}, A.~W. 2011,
  Physical Review Letters, 106, 081101, 1011.6142

\bibitem[{{Papaloizou} \& {Pringle}(1978)}]{1978MNRAS.182..423P}
{Papaloizou}, J., \& {Pringle}, J.~E. 1978, \mnras, 182, 423

\bibitem[{Passamonti {et~al.}(2008)Passamonti, Stavridis, \&
  Kokkotas}]{PhysRevD.77.024029}
Passamonti, A., Stavridis, A., \& Kokkotas, K.~D. 2008, Phys. Rev. D, 77,
  024029

\bibitem[{{Patruno} \& {Watts}(2012)}]{2012arXiv1206.2727P}
{Patruno}, A., \& {Watts}, A.~L. 2012, ArXiv e-prints, 1206.2727

\bibitem[{{Pons} {et~al.}(2005){Pons}, {Gualtieri}, {Miralles}, \&
  {Ferrari}}]{2005MNRAS.363..121P}
{Pons}, J.~A., {Gualtieri}, L., {Miralles}, J.~A., \& {Ferrari}, V. 2005,
  \mnras, 363, 121, arXiv:astro-ph/0504062

\bibitem[{{Rezania} \& {Morsink}(2002){Rezania} \& {Morsink}}]{2002ApJ...574..908M}
{Rezania}, V. \& Morsink, S. M. 2002,
\apj, 574, 908, arXiv:astro-ph/0111571

\bibitem[{{Rezzolla} {et~al.}(2000){Rezzolla}, {Lamb}, \& {Shapiro}}]{Luciano1}
{Rezzolla}, L., {Lamb}, F., \& {Shapiro}, L.~S. 2000,
  \apjl, 531, L141, arXiv:astro-ph/9911188
  
  \bibitem[{{Rezzolla} {et~al.}(2001a){Rezzolla}, {Lamb}, \& {Shapiro}}]{Luciano2}
{Rezzolla}, L., {Lamb}, {Markovic}, D.,  F., \& {Shapiro}, L.~S. 2000,
  \prd, 64, 104013, arXiv:astro-ph//010762

\bibitem[{{Rezzolla} {et~al.}(2001b){Rezzolla}, {Lamb}, \& {Shapiro}}]{Luciano3}
{Rezzolla}, L., {Lamb}, F., {Markovic}, D., \& {Shapiro}, L.~S. 2000,
  \prd, 64, 104014, arXiv:astro-ph/010761

\bibitem[{{Saio}(1982)}]{1982ApJ...256..717S}
{Saio}, H. 1982, \apj, 256, 717

\bibitem[{{Schenk} {et~al.}(2002){Schenk}, {Arras}, {Flanagan}, {Teukolsky}, \&
  {Wasserman}}]{2002PhRvD..65b4001S}
{Schenk}, A.~K., {Arras}, P., {Flanagan}, {\'E}.~{\'E}., {Teukolsky}, S.~A., \&
  {Wasserman}, I. 2002, \prd, 65, 024001, arXiv:gr-qc/0101092

\bibitem[{{Shibazaki} {et~al.}(1989){Shibazaki}, {Murakami}, {Shaham}, \&
  {Nomoto}}]{1989Natur.342..656S}
{Shibazaki}, N., {Murakami}, T., {Shaham}, J., \& {Nomoto}, K. 1989, \nat, 342,
  656

\bibitem[{{Shternin} \& {Yakovlev}(2008)}]{2008PhRvD..78f3006S}
{Shternin}, P.~S., \& {Yakovlev}, D.~G. 2008, \prd, 78, 063006, 0808.2018

\bibitem[{{Shternin} {et~al.}(2011){Shternin}, {Yakovlev}, {Heinke}, {Ho}, \&
  {Patnaude}}]{2011MNRAS.412L.108S}
{Shternin}, P.~S., {Yakovlev}, D.~G., {Heinke}, C.~O., {Ho}, W.~C.~G., \&
  {Patnaude}, D.~J. 2011, \mnras, 412, L108, 1012.0045

\bibitem[{Villain {et~al.}(2005)Villain, Bonazzola, \&
  Haensel}]{PhysRevD.71.083001}
Villain, L., Bonazzola, S., \& Haensel, P. 2005, Phys. Rev. D, 71, 083001

\bibitem[{{Wang} {et~al.}(2011){Wang}, {Zhang}, {Zhao}, {Kojima}, {Yin}, \&
  {Song}}]{2011A&A...526A..88W}
{Wang}, J., {Zhang}, C.~M., {Zhao}, Y.~H., {Kojima}, Y., {Yin}, H.~X., \&
  {Song}, L.~M. 2011, \aap, 526, A88, 1011.5013

\bibitem[{{Watts}(2012)}]{2012ARA&A..50..609W}
{Watts}, A.~L. 2012, \araa, 50, 609, 1203.2065

\bibitem[{{Yakovlev} {et~al.}(2008){Yakovlev}, {Gnedin}, {Kaminker}, \&
  {Potekhin}}]{2008AIPC..983..379Y}
{Yakovlev}, D.~G., {Gnedin}, O.~Y., {Kaminker}, A.~D., \& {Potekhin}, A.~Y.
  2008, in American Institute of Physics Conference Series, Vol. 983, 40 Years
  of Pulsars: Millisecond Pulsars, Magnetars and More, ed. {C.~Bassa, Z.~Wang,
  A.~Cumming, \& V.~M.~Kaspi}, 379--387

\bibitem[{{Yakovlev} {et~al.}(1999){Yakovlev}, {Kaminker}, \&
  {Levenfish}}]{1999A&A...343..650Y}
{Yakovlev}, D.~G., {Kaminker}, A.~D., \& {Levenfish}, K.~P. 1999, \aap, 343,
  650, arXiv:astro-ph/9812366

\bibitem[{{Yakovlev} \& {Pethick}(2004)}]{2004ARA&A..42..169Y}
{Yakovlev}, D.~G., \& {Pethick}, C.~J. 2004, \araa, 42, 169,
  arXiv:astro-ph/0402143

\bibitem[{{Yoshida} \& {Lee}(2000{\natexlab{a}})}]{2000ApJ...529..997Y}
{Yoshida}, S., \& {Lee}, U. 2000{\natexlab{a}}, \apj, 529, 997,
  arXiv:astro-ph/9908197

\bibitem[{{Yoshida} \& {Lee}(2000{\natexlab{b}})}]{2000ApJS..129..353Y}
------. 2000{\natexlab{b}}, \apjs, 129, 353, arXiv:astro-ph/0002300

\bibitem[{{Yoshida} \& {Lee}(2001)}]{2001ApJ...546.1121Y}
------. 2001, \apj, 546, 1121, arXiv:astro-ph/0006107

\bibitem[{{Zhang} \& {Kojima}(2006)}]{2006MNRAS.366..137Z}
{Zhang}, C.~M., \& {Kojima}, Y. 2006, \mnras, 366, 137, arXiv:astro-ph/0410248

\end{thebibliography}
\end{document}